\documentclass[10pt,aps,pra,twocolumn,superscriptaddress,longbibliography,nobalancelastpage,floatfix]{revtex4-2}
\usepackage[T1]{fontenc}
\usepackage[utf8]{inputenc}
\newcommand{\apptoctype}{toc}
\pdfoutput=1 
\usepackage{graphicx}
\usepackage{mathtools}
\usepackage[explicit]{titlesec} % Supplement titles
\usepackage{bm}
\usepackage{physics}
\usepackage[colorlinks = true,
            allcolors = {Blue}]{hyperref}
\usepackage[capitalize]{cleveref}
\usepackage[caption=false]{subfig}
\usepackage[dvipsnames]{xcolor}

\usepackage{amssymb}

\usepackage{enumitem}
\setlist[itemize]{label=\textbullet}

\usepackage[normalem]{ulem}

\begin{document}

\author{Qingyi Zhou}
\affiliation{Department of Electrical and Computer Engineering,
University of Wisconsin-Madison, Madison, WI 53706, USA}
\author{Piper Fowler-Wright}
\affiliation{Department of Chemistry and Biochemistry, University of California San Diego, La Jolla, CA 92093, USA}
\author{Zongfu Yu}
\affiliation{Department of Electrical and Computer Engineering,
University of Wisconsin-Madison, Madison, WI 53706, USA}
\author{Joel Yuen-Zhou}
\affiliation{Department of Chemistry and Biochemistry, University of California San Diego, La Jolla, CA 92093, USA}
\author{Michael Reitz}
\affiliation{Department of Chemistry and Biochemistry, University of California San Diego, La Jolla, CA 92093, USA}

\title{Full-wave nonlinear microscopy reveals guided channel for ultrafast polariton transport}
% Previous title: Nonlinear spectroscopy of ultrafast polariton transport in arbitrary electromagnetic environments via FDTD simulations
 
\date{\today} 
\begin{abstract} 

%\vspace{0.8\baselineskip}\noindent\small{DOI: \href{}{} }

% Piper's previous version 
% We show how ultrafast nonlinear spectroscopy can be readily integrated within finite-difference time-domain (FDTD) simulations, enabling the modeling of pump–probe experiments in arbitrary electromagnetic environments. 
% We focus on polariton transport in strongly coupled light–matter systems, establishing a perturbative framework to study the ultrafast propagation of hybrid light–matter excitations in complex nanophotonic structures. 
% We benchmark the method against quantum-optical predictions based on Tavis–Cummings models for simple cavity-emitter systems, before applying it to realistic electromagnetic environments exhibiting non-trivial spatial structure and supporting multiple mode families. 
% As a novel application, we use inverse-design techniques to realize a mode converter that transforms radiative cavity polaritons into guided polaritons, increasing their propagation length by one order of magnitude, before converting them back into radiative modes, enabling pump--probe readout of the transported excitation. 
% Our method enables a self-consistent modeling of ultrafast spectroscopy in complex electromagnetic environments where the back-action of the induced polarization on the light field is important, providing a versatile platform for studying polariton dynamics in realistic nanophotonic geometries beyond few-mode or few-emitter approximations. 

% New version
We show how finite-difference time-domain (FDTD) simulations can be extended to model ultrafast nonlinear microscopy, enabling the prediction of spatially-resolved pump–probe signals in arbitrary electromagnetic environments.
Focusing on polariton transport in strongly coupled light–matter systems, we develop a perturbative framework to study the ultrafast propagation of hybrid light–matter excitations in nanophotonic structures.
% After benchmarking against Tavis–Cummings predictions for simple cavity-emitter systems, we apply the framework to a distributed Bragg reflector (DBR) cavity and show that the structure supports radiatively protected guided modes below the light line, which are typically neglected in single-mode-family descriptions. 
We first apply the framework to a standard distributed Bragg reflector (DBR) cavity, reproducing established results for polariton transport from a multimode Tavis--Cummings model. 
We then consider the full modal landscape of the same cavity, including guided modes below the light line that are typically neglected in single-mode-family descriptions. 
Exploiting these modes, we design compact mode converters that transfer radiative cavity polaritons into photon-like guided polaritons and back, utilizing the guided modes for low-loss propagation. 
Despite molecular dephasing, this enables transport of molecular excitation over a millimeter, an order of magnitude beyond current transport experiments. 
We further compute the pump–probe differential transmission signal, providing an experimental signature of the mechanism. 
% In all, our results show that the full modal landscape of a photonic structure can be utilized to shape transport in polaritonic systems. 
% With full-wave modeling, this landscape can be engineered to bypass constraints commonly assumed to be intrinsic to molecular polaritons. 
Our results show that the modal landscape of a photonic cavity can be engineered to bypass limitations commonly assumed to be intrinsic to the transport of molecular polaritons. 
\end{abstract}

\maketitle

\section{Introduction} % The new version
Exciton polaritons are hybrid light-matter quasiparticles formed through strong coupling of confined electromagnetic modes with electronic excitations. 
They have gained significant attention over the past decade, due to their potential applications in areas such as polariton chemistry \cite{ribeiro2018}, condensation \cite{keeling2020, byrnes2014exciton, lerario2017room} and lasing \cite{strashko2018}, as well as charge \cite{orgiu2015conductivity, hagenmuller2017cavity} and energy transport \cite{sandik2024, hou2020, myers2018polariton, schachenmayer2015cavity, feist2015extraordinary}, where the latter may enable novel optoelectronic platforms \cite{lerario2016,hou2020,liu2023}.
% More specific background related to transport
Specifically, a central bottleneck in traditional organic optoelectronic platforms is the short-range transport of molecular excitations, typically limited to a few tens of nm \cite{mikhnenko2015exciton, kohler2015electronic}. Exciton polaritons are predicted to overcome this and allow for fast and long-range excitation transport \cite{lerario2016, hou2020, liu2023, liu2023long, sandik2024}. 
Further, recent advances in ultrafast nonlinear microscopy have enabled the imaging of molecular polariton transport on picosecond timescales, revealing unexpected sub-group-velocity transport as well as a crossover from ballistic to diffusive propagation behavior~\cite{xu2023,balasubrahmaniyam2023,rozenman2018,pandya2021,pandya2022,jin2023,berghuis2022controlling}. 
Theoretical studies have identified scattering of polaritons into ``dark excitonic states'' (i.e., the molecular absorption band) as a key mechanism responsible for slowing down polariton transport and altering the propagation characteristics~\cite{groenhof2019,sokolovskii2023,tichauer2023, Suyabatmaz2023vibrational, engelhardt2023polariton,Ying2024, chng2025,krupp2024,khazanov2023,fowlerwright2026mapping}. 
% These results have prompted extensive research \cite{} into the mechanism of transport of hybrid light-matter states in such systems. Despite of these works, significant challenges and open questions remain in understanding the nonlinear polariton response as observed in pump-probe experiments. 

\begin{figure*}[t]
    \centering
    \includegraphics[width=1.0\textwidth]{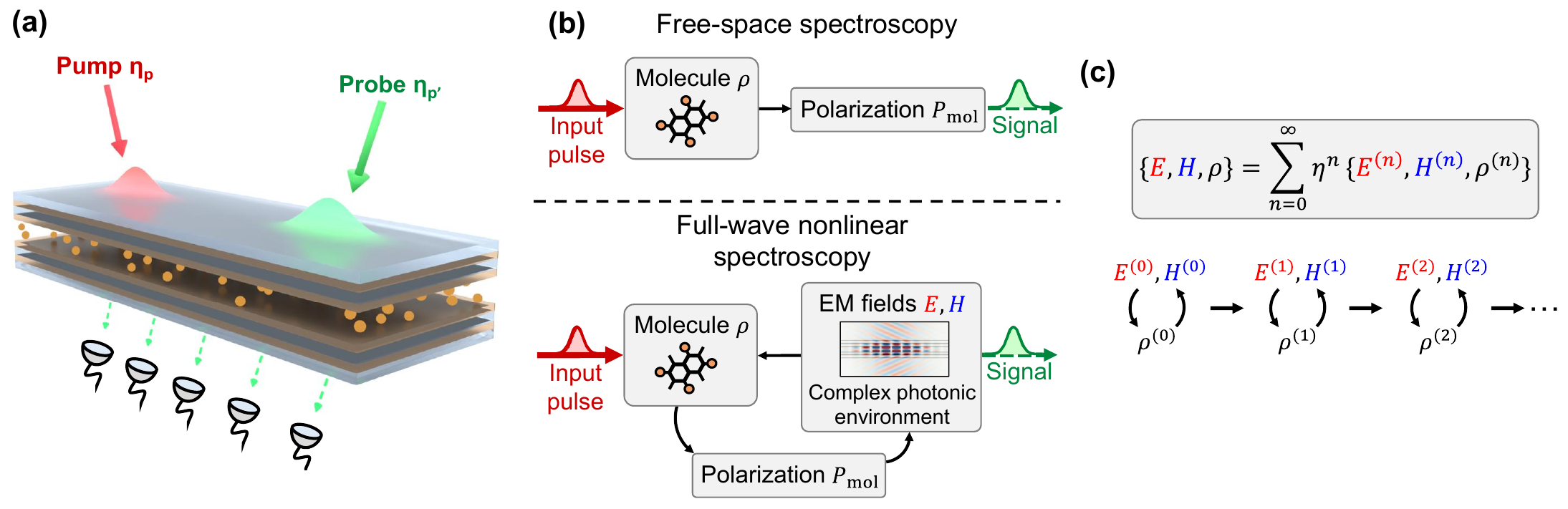}
    \caption{Full-wave nonlinear spectroscopy. 
    (a) Prototypical setup for spatially resolved pump--probe spectroscopy with light-matter back-action: an ensemble of molecules is placed inside an optical cavity formed by DBR mirrors, which is driven by input pulses with amplitudes $\eta_{p}$ (pump) and $\eta_{p'}$ (probe). 
    (b) Schematic comparison between the proposed ``full-wave nonlinear spectroscopy'' and conventional ``free-space spectroscopy''. In the conventional framework, molecules are driven by a prescribed incident field. In contrast, our framework accounts for the feedback of the molecular response onto the optical field, enabling a self-consistent treatment of coupled light--matter dynamics in complex and structured electromagnetic environments. 
    (c) Hierarchy of the perturbative expansion. The nonlinear electromagnetic fields and molecular density matrices are constructed iteratively from the lower-order light--matter dynamics. }
    \label{fig:fig1}
\end{figure*}

% In most existing studies, all stages of excitation transport, including injection, propagation, and readout, take place within a single polariton branch \cite{groenhof2019, sokolovskii2023, tichauer2023, Ying2024, krupp2024, khazanov2023, fowlerwright2026mapping}. 
These studies reveal a trade-off in polariton transport: photon-like polaritons propagate efficiently but carry less molecular excitation, whereas exciton-like polaritons are vulnerable to dephasing ~\cite{groenhof2019,sokolovskii2023,tichauer2023, Suyabatmaz2023vibrational, engelhardt2023polariton,Ying2024, chng2025,krupp2024,khazanov2023,fowlerwright2026mapping}. 
% Under this single-branch picture, a trade-off emerges: photon-like polaritons propagate efficiently but carry less molecular excitation, whereas exciton-like polaritons are vulnerable to dephasing. 
In this work, we show that this trade-off is not fundamental but may be overcome by separating the polariton channels used for excitation or readout from those used for propagation. 
This strategy is naturally enabled by realistic photonic structures, which typically support many families of optical modes.
In particular, the distributed Bragg reflector (DBR) structures commonly used in polariton experiments support not only radiative cavity modes, but also guided modes below the light cone, which can serve as low-loss transport channels \cite{liu2023, liu2023long, hou2020}. 
By assigning injection and readout to externally addressable cavity modes, and long-range propagation to radiatively protected guided modes, the trade-off can be circumvented. 

Verifying a multimode transport mechanism requires connecting it to experimentally accessible pump--probe microscopy signals.
While the perturbative formalism of nonlinear spectroscopy is well established in free-space \cite{mukamel1995principles}, the computation and interpretation of pump--probe signals in confined photonic environments poses a significant challenge, since the molecular polarization feeds back onto, and thereby reshapes, the electromagnetic field itself \cite{f2018theory}. 
% Existing works / approaches
Recent works have provided perturbative theories of polariton spectroscopy for single-mode \cite{reitz2025nonlinear, reitz2026multidimensional} and multimode planar cavities \cite{fowlerwright2026mapping}, yet they rely on simplified cavity descriptions that do not generalize directly to realistic and complex environments, such as the multiple mode families supported by DBR cavities.
Conversely, full-wave approaches based on Maxwell–Liouville equations naturally handle arbitrary geometries \cite{ziolkowski1995ultrafast, zhou2024, zhou2024b, dini2024nonlinear, chen2019, bustamante2025molecular, sidler2025density, Ji2026maxwelllink}, yet do not disentangle the nonlinear response into spectroscopically meaningful orders. What is needed, therefore, is a unified framework that preserves the full electrodynamic complexity of Maxwell-based simulations while allowing nonlinear observables such as differential pump–probe spectra to be computed and decomposed into their underlying pathways. % electromagnetic realism with spectroscopic interpretability

% Summarize the paper: our contribution
In this work, we address the above challenges by introducing a framework for modeling ultrafast nonlinear microscopy in arbitrary photonic environments. To this end, we combine perturbative nonlinear spectroscopy with semiclassical Maxwell–Liouville equations that can be solved efficiently using full-wave finite-difference time-domain (FDTD) simulations \cite{taflove2005computational}, thereby extending earlier approaches \cite{reitz2025nonlinear, fowlerwright2026mapping} to complex nanophotonic structures. 
We first apply the framework to a standard DBR cavity, reproducing established behavior including ballistic polariton transport and dephasing-induced velocity renormalization. 
We then uncover the full modal landscape of the same DBR structure, identifying guided modes below the light cone that are nearly lossless and largely immune to molecular dephasing. 
We show how to access these modes by inversely designing mode converters that transform between radiative and guided mode families \cite{Molesky2018, lalau2013adjoint}. 
By assigning complementary roles to distinct mode families, we demonstrate molecular excitation transport over $1$\,mm in the presence of molecular dephasing, an order of magnitude beyond state-of-the-art results \cite{hou2020, balasubrahmaniyam2023, rozenman2018, pandya2021,liu2023,berghuis2022controlling, liu2023long, Xie20252dmaterial, xu2023, dang2024long, wurdack2021motional}. 
Finally, we calculate the pump--probe differential transmission signal at the remote port, providing a spectroscopic prediction directly applicable to experiment.

\section{Perturbative Maxwell-Liouville approach }

We start by considering $N$ molecules located at positions $\bm{r}_j$ ($j\in \{ 1,\hdots, N\}$) inside an arbitrary photonic environment. The molecules are treated as identical two-level systems (TLSs) with transition frequency $\omega_{0}$ and dipole operator $\hat{\bm{\mu}}_{j} = \bm{\mu} \bigl( \hat{\sigma}_{j}^{-}  + \hat{\sigma}_{j}^{+} \bigr)$. 
Note that we use $\hat{\sigma}_{j}^{\pm}$ and $\hat{\sigma}_{j}^{z}$ to denote the standard Pauli raising/lowering and inversion operators, respectively. 
While more elaborate emitter models can be incorporated, including additional internal states, vibronic structure, energetic disorder, or non-Markovian vibronic dynamics \cite{del2018tensor, fowler2022efficient, schwennicke2024extracting}, such effects can also modify polariton velocity renormalization \cite{xu2023, Ying2024, krupp2024}. Here, we adopt a minimal two-level description that keeps the formalism transparent while capturing the essential light-mediated interactions and transport physics. 
A typical pump-probe setup involving multiple molecules inside an optical cavity is shown schematically in Fig.~\ref{fig:fig1}(a). % Schematics 
The mean-field Hamiltonian of such TLSs coupled to the electromagnetic fields can be written as
\begin{align}
H_\text{MF}(t) &= H_{0} + H_\text{int}(t) = \sum_{j=1}^{N} H_{0}^{j} + \sum_{j=1}^{N} H^{j}_\text{int}(t) \nonumber \\
    &=\sum_{j=1}^N \frac{\hbar\omega_0}{2} \hat{\sigma}_{j}^{z}
  - \sum_{j=1}^N \bm{\mu} \cdot\bm E(\bm{r}_{j}, t)\,\bigl( \hat{\sigma}_{j}^{-}  + \hat{\sigma}_{j}^{+} \bigr),
\label{eq:H_MF}
\end{align}
where $H_0^j$ describes the free molecular Hamiltonian and $H_\text{int}^j$ accounts for the dipole coupling between the molecules and the local electric field $\bm E(\bm{r}_j)$. Here, ``mean-field'' refers to the factorization between light and matter degrees of freedom that allows for the electromagnetic field  to be treated classically, while the molecules remain quantum degrees of freedom.
% Without loss of generality, all molecular dipole moments are aligned along the $x$-direction. 
The density matrix $\rho_{j}$ of the $j$-th molecule evolves according to the quantum master equation
\begin{equation}
\frac{d\rho_{j}}{dt} 
  = -\frac{i}{\hbar}\bigl[H_{j}(t),\,\rho_{j}\bigr] + \mathcal{D}[\rho_{j}],
\label{eq:master}
\end{equation}
where $H_{j}(t) = H_{0}^{j} + H^{j}_\text{int}(t) $ denotes the single-molecule Hamiltonian, and the dissipator $\mathcal{D}[\rho_{j}]$ represents the Lindbladian terms that incorporate additional loss channels of the molecule; below we consider pure dephasing at rate $\gamma_{\phi}$ \cite{fowlerwright2026mapping}. 

To model the spectroscopic experiment, we start by assuming a single input field with amplitude $\eta_{p}$ \cite{reitz2025nonlinear}. 
The evolution of the classical electromagnetic fields is governed by Maxwell's equations: 
\begin{equation}
    \begin{aligned}
    \nabla \times \bm{E} &= -\mu_{0}\partial_{t}\bm{H} - \eta_{p} \bm{M}_{\mathrm{in}}, \\ 
    \nabla \times \bm{H} &= \varepsilon_{r}\varepsilon_{0}
      \partial_{t}\bm{E}
      + \bm{J}_{\mathrm{mol}} + \eta_{p}  \bm{J}_{\mathrm{in}},
    \label{eq:maxwell}
    \end{aligned}
\end{equation}
where $\bm{J}_{\mathrm{in}}$ and $\bm{M}_{\mathrm{in}}$ correspond to the normalized spatiotemporal profiles of the external current sources used to generate the input pulse, while the parameter $\eta_{p}$ controls the amplitude and serves as the expansion parameter below. 
% On the other hand, the current $\bm J_{\mathrm{mol}}(\bm r, t) = \sum_{j} \partial_{t}\Tr\!\bigl[\hat{\bm{\mu}}_{j} {\rho}_{j}(t)\bigr]\,\delta(\bm{r} - \bm{r}_{j})$ arises from the dipole moment of each TLS and describes the feedback onto the classical electromagnetic field, with $\delta(\bm{r})$ being the Dirac $\delta$-function. % \partial_{t}\bm{P}^{\mathrm{mol}}
The molecular polarization is $\bm P_{\mathrm{mol}}(\bm r,t)=\sum_j\Tr\!\bigl[\hat{\bm\mu}_j\rho_j(t)\bigr]\delta(\bm r-\bm r_j)$, and the corresponding current $\bm J_{\mathrm{mol}}=\partial_t\bm P_{\mathrm{mol}}$ describes the feedback onto the classical electromagnetic field, with $\delta(\bm r)$ being the Dirac $\delta$-function. 
A detailed derivation can be found in the Supplemental Material (SM)~\cite{sm}. 
The above coupled Eqs.~\eqref{eq:master} and \eqref{eq:maxwell} form the self-consistent Maxwell--Liouville framework, which can be time-evolved via FDTD~\cite{zhou2024}. By contrast, the previous works Refs.~\cite{reitz2025nonlinear, fowlerwright2026mapping, reitz2026multidimensional} assumed a simplified evolution of Eqs.~\eqref{eq:maxwell} in which the photonic modes are already prescribed. % (details can be found in the SM \cite{sm})

\begin{figure*}[t]
    \centering
    \includegraphics[width=\textwidth]{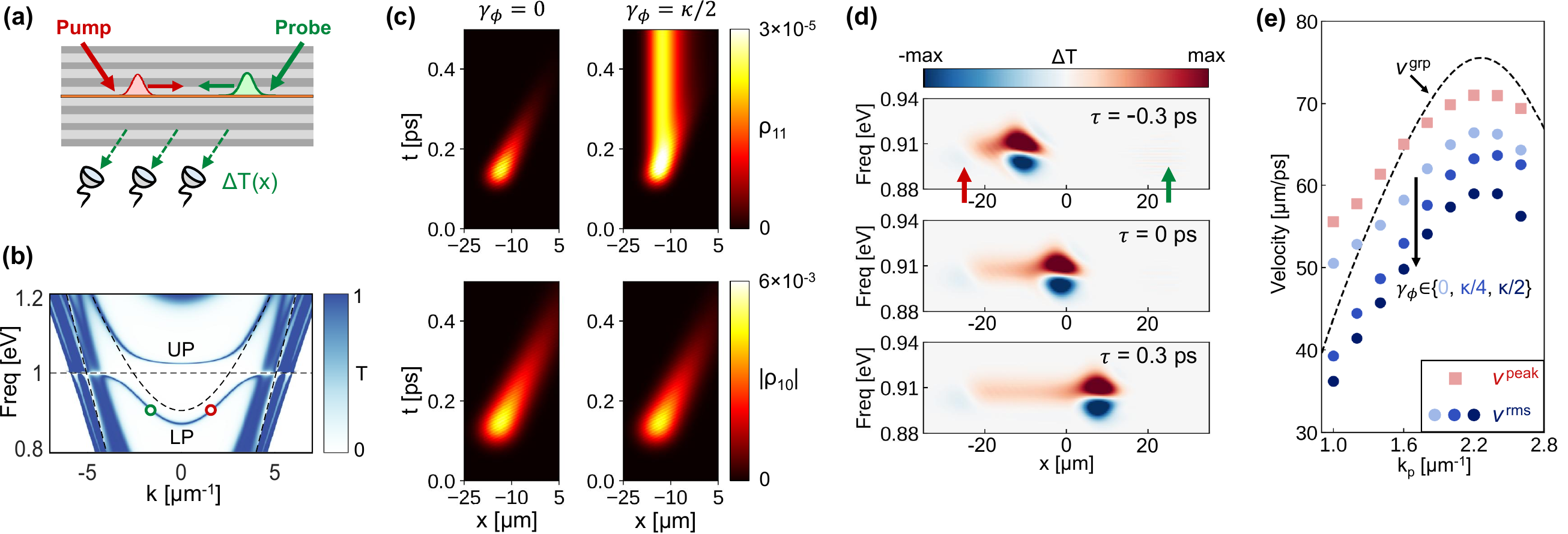}
    \caption{Full-wave pump-probe microscopy results for molecules inside a DBR cavity. 
    (a) Schematics of the pump-probe experiment, where a planar DBR cavity (gray) containing a molecular layer (orange) is driven by counter-propagating pump ($p$) and probe ($p'$) pulses. Information about pump-induced transport is retrieved through the differential signal $\Delta T(x)$, detected along the probe direction. 
    (b) TMM results of the dispersion relation, revealing the UP and LP branches. The red (green) dot indicates the in-plane wave vector of the incident pump (probe) pulse, centered at $k_{p} = -k_{p'} = \pi/2~\mu\text{m}^{-1}$. The molecules' transition frequency is fixed as $\omega_{0} = 1~\text{eV}$. The dashed lines indicate the molecular transition frequency, the bare cavity dispersion, and the light line.
    (c) Pump-only dynamics for both molecular population $\rho_{11}$ and coherence $|\rho_{10}|$. 
    In the absence of dephasing, both $\rho_{11}$ and $|\rho_{10}|$ show ballistic transport; with dephasing $\gamma_{\phi}=\kappa/2$ a static population emerges since bright states are transferred into immobile dark states. 
    (d) Spatiotemporal pump-probe spectra $\Delta T(x, \omega)$ for different time delays $\tau\in \{ -0.3, 0, 0.3\}~\mathrm{ps}$. In the top panel, arrows indicate the spatial positions of the pump (red) and probe (green) injection positions. 
    (e) Transport velocity of the circular feature ($v_{k_{p}}^\text{peak}$) and rms displacement ($v_{k_{p}}^\text{rms}$) for various pump wave vectors $k_{p}$. Larger $\gamma_{\phi}$ leads to stronger group velocity renormalization. 
    }
    \label{fig:fig2}
\end{figure*}
Fig.~\ref{fig:fig1}(b) compares our approach, which we refer to as ``full-wave nonlinear spectroscopy'', with conventional free-space spectroscopy. %where the backaction of the polarization onto the field can typically be neglected. 
Our framework treats the coupling between electromagnetic fields and matter in a self-consistent manner by taking into account the backaction of the molecular polarization onto the electromagnetic field. 
% Perturbation expansion: pump-only
Although the above set of mean-field equations can be solved numerically \cite{sukharev2011numerical, carollo2021, zhou2024, dini2024nonlinear}, this does not, by itself, provide transparent insight into the structure of the nonlinear response. 
To understand the nonlinear response in a systematic way, we follow the standard approach of nonlinear spectroscopy \cite{mukamel1995principles} and expand both the electromagnetic fields and the matter density matrix perturbatively in powers of the input-field amplitude $\eta_{p}$ \cite{reitz2025nonlinear, fowlerwright2026mapping}:
\begin{equation}
\begin{aligned}
    \{ \bm{E}, \bm{H} \} &= \sum_{n=0}^{\infty}
    \eta_p^{\,n}\cdot \{ \bm{E}^{(n)}, \bm{H}^{(n)} \},\\
    {\rho}_{j} &= \sum_{n=0}^{\infty}
    \eta_p^{\,n} \cdot {\rho}_{j}^{(n)},
\end{aligned}
\label{eq:pump_expansion}
\end{equation}
which lead to
\begin{equation}
    \begin{aligned}
    \nabla \times \bm{E}^{(n)} &= -\mu_{0}\partial_{t}\bm{H}^{(n)} - \delta_{1n} \bm{M}_{\mathrm{in}}, \\ 
    \nabla \times \bm{H}^{(n)} &= \varepsilon_{r}\varepsilon_{0}
      \partial_{t}\bm{E}^{(n)}
      + \bm{J}^{(n)}_{\mathrm{mol}} +\delta_{1n} \bm{J}_{\mathrm{in}},
    \label{eq:maxwell_expansion}
    \end{aligned}
\end{equation}
\begin{equation}
\begin{aligned}
\frac{d\rho_{j}^{(n)}}{dt}
&= -\frac{i}{\hbar}\Bigl\{
   [H_{0}^{j},\rho_{j}^{(n)}] \\
&\quad - \sum_{i=0}^{n}
   \bigl[\hat{\bm{\mu}}_{j} \cdot \bm{E}^{(n-i)}(\bm{r}_{j}, t),\,\rho_{j}^{(i)}\bigr]
\Bigr\}
+ \mathcal{D}[\rho_{j}^{(n)}].
\end{aligned}
\end{equation}
The above equations can be viewed as a perturbative decomposition of the well-known Maxwell-Liouville equations. The term $\bigl[\hat{\bm\mu}_{j} \cdot \bm{E}^{(n-i)},\rho^{(i)}_{j} \bigr]$ couples $\rho_{j}^{(n)}$ to lower-order quantities, so that the response at each perturbative order is driven by the dynamics obtained at preceding orders.
Equations at different order $n$ can therefore be viewed as a set of coupled PDEs which are solved by running multiple FDTDs concurrently. 
A simplified perturbation hierarchy showing the iterative construction of the nonlinear response is shown in Fig.~\ref{fig:fig1}(c). 

\section{Polariton transport in DBR cavity}

Having established the perturbative Maxwell--Liouville framework, we begin by examining polariton transport in a standard DBR cavity. 
This regime has been studied extensively \cite{reitz2025nonlinear,fowlerwright2026mapping, sokolovskii2023, tichauer2023, Ying2024} and provides the baseline for the long-range transport we demonstrate later.
The setup is shown in Fig.~\ref{fig:fig2}(a), where we consider a layered structure with bare cavity frequency $\omega_{c}=0.9~\text{eV}$ (further details are provided in the SM~\cite{sm}). 
A molecular layer is placed at the center of the cavity and couples strongly to the cavity mode which has a photon loss rate of %denoted as $\kappa$, the value is 
$\kappa\approx 0.01$ eV. 
By calculating the linear transmission of this system using the transfer-matrix method (TMM), the dispersion relations of the two bright eigenstates, namely the lower polariton (LP) and the upper polariton (UP), can be extracted, as shown in Fig.~\ref{fig:fig2}(b). 
Here we consider a resonant excitation scheme, matching the one used in Ref.~\cite{fowlerwright2026mapping}: a pump pulse ($p$) resonantly excites a certain mode on the LP branch (red dot in Fig.~\ref{fig:fig2}(b)) with $k_{p}>0$, while a probe pulse ($p'$) with $k_{p'}<0$ is injected and counter-propagates. 
The temporal dependence of the input fields are proportional to $\eta_{j} f_{j}(t-\tau_{j}) e^{-i \omega_{j} t} $ for $j\in \{ p, p'\}$, where $\eta_{j}$, $f_{j}$, $\omega_{j}$ and $\tau_{j}$ denote the amplitude, normalized envelope, carrier frequency, and arrival time of the pump and probe pulses, respectively. 
In the following, we assume the pulses to have Gaussian temporal and spatial profiles with temporal width $\sigma_{t} = 25~\text{fs}$ and spatial width $\sigma_{r} = 5~\mu\text{m}$. 
% While the polariton transport inside such system is well understood \cite{}, this section serves as a benchmark, which shows that our full-wave simulation produces known results. 

%\subsection{Pump-only dynamics}
Before analyzing the pump-probe experiments, we examine the pump-only dynamics by setting $\eta_{p'}=0$. % spectroscopic signal observed
A pump pulse resonantly excites the LP branch at $k_{p} = \pi/2~\mu\text{m}^{-1}$. 
Fig.~\ref{fig:fig2}(c) displays the spatio-temporal evolution of both the molecular population $\rho_{11}(x, t)$ and coherence $\rho_{10}(x, t)$, for molecular dephasing $\gamma_{\phi}=0$ and $\gamma_{\phi}=\kappa/2$. 
In the absence of dephasing, both $\rho_{11}$ and $\rho_{10}$ propagate at the same group velocity, revealing ballistic excitation transport. 
Introducing dephasing significantly alters the dynamics: while the coherence $\rho_{10}$ simply decays more rapidly, the population $\rho_{11}$ develops a ``dragged'' trail from the passing polariton wave packet. This behavior arises because pure dephasing $\gamma_{\phi}>0$ transfers population from the coherent bright state into incoherent, immobile dark states. 
Local incoherent excitons are created but remain stationary, which makes the excitation transport much slower than the idealized ballistic transport \cite{groenhof2019, sokolovskii2023, Ying2024, fowlerwright2026mapping}. 

We next include both pump $p$ and probe $p'$. 
The pump--probe delay is defined from the pulse injection times as $\tau = \tau_{p'} - \tau_{p}$. 
% The $H_{z}$ field for special case $\tau=0$ is visualized in Fig.~\ref{fig:fig3} to help understanding?
Following the framework proposed in \cite{reitz2025nonlinear, fowlerwright2026mapping}, we expand all field quantities in pump amplitude $\eta_{p}$ and probe amplitude $\eta_{p'}$:
\begin{equation}
\begin{aligned}
    \{ \bm{E}, \bm{H} \} &= \sum_{n=0}^{\infty}\sum_{n'=0}^{\infty}
    \eta_p^{\,n}\eta_{p'}^{\,n'}\cdot \{ \bm{E}^{(n)(n')}, \bm{H}^{(n)(n')} \},\\
    {\rho}_{j} &= \sum_{n=0}^{\infty}\sum_{n'=0}^{\infty}
    \eta_p^{\,n}\eta_{p'}^{\,n'} \cdot {\rho}_{j}^{(n)(n')},
\end{aligned}
\label{eq:pp_expansion}
\end{equation}
where the indices $n$ and $n'$ refer to the orders in the pump and probe fields, respectively. 
By substituting the above expressions into the mean-field equations, a hierarchical system of equations describing the pump-probe spectroscopy can be derived~\cite{sm}. 
Representative signals from different $(n)(n')$ orders are provided alongside the full pump-probe equations in the SM~\cite{sm}. 

In pump-probe experiments, the observable related to nonlinear effects typically comes from the differential signal between
pump-on and pump-off conditions. 
Here we consider the differential transmission measured at the output monitor plane, $\Delta T(\omega)=T^{\mathrm{pump\,on}}_{p'}(\omega)-T^{\mathrm{pump\,off}}_{p'}(\omega)$. 
Similar to the definition in Ref.~\cite{fowlerwright2026mapping}, the lowest-order contribution to the position-dependent differential transmission can be derived as
\begin{equation}
\begin{aligned}
    \Delta T(\bm{r}, \omega)
    &\propto -\frac{\eta_{p}^{2}}{2}
    \operatorname{Re}\Big\{
    E_{x}^{(0)(1)}(\bm{r}, \omega)\,
    H_{z}^{(2)(1)*}(\bm{r}, \omega) \\
    &\qquad\qquad
    + E_{x}^{(2)(1)}(\bm{r}, \omega)\,
    H_{z}^{(0)(1)*}(\bm{r}, \omega) 
    \Big\},
\end{aligned}
\end{equation}
scaling with the intensity of the pump field.
Note that, in addition, a $k$-space filtering is carried out along the probe direction, which helps to separate the pump--probe response from undesired double-quantum coherence contributions~\cite{sm, Hamm_Zanni_2011}. 
% Poynting vector: Sy ~ 1/2 Re[-Ex Hz*]
In Fig.~\ref{fig:fig2}(d), we visualize the calculated signal $\Delta T(\bm{r}, \omega)$ following pump and probe pulses with opposite in-plane wave vectors $k_{p}=-k_{p'}=\pi/2~\mu\text{m}^{-1}$. % centered at $x=-25~\mu\text{m}$ and $x=25~\mu\text{m}$
Here the dephasing is fixed as $\gamma_{\phi} = \kappa/2$. 
% The setup is similar to that in Ref.~\cite{fowlerwright2026mapping} since we are reproducing the existing results. 
Three different probe delays $\tau \in \{ -0.3, 0, 0.3\}~\text{ps}$ are examined. Negative delays are included since the probe pulse must first propagate inside the structure before reaching the region where it spatially overlaps with the pump pulse. 
The leading circular feature in $\Delta T(\bm{r}, \omega)$ shows red-blue lobes, resulting from Rabi contraction following blue shift of the LP \cite{delpo2020, reitz2025nonlinear}. 
In addition, due to $\gamma_{\phi}>0$, a trailing feature emerges in $\Delta T(\bm{r}, \omega)$, corresponding to the population of static dark excitonic states. 

Finally, based on the simulated nonlinear microscopy results, we obtain the peak and root-mean-square (rms) displacement of the $| \Delta T(\bm{r}, \omega)|$ profile, then extract the corresponding velocities $v_\text{peak}$ and $v_\text{rms}$ through linear fitting (see SM~\cite{sm} for details). 
We repeat the procedure for different in-plane wave vectors $|k_{p}| = |k_{p'}|$. The extracted group velocities $v_\text{peak}$ and $v_\text{rms}$ are displayed in Fig.~\ref{fig:fig2}(e). % extracted group velocity put in fig2e
It can be concluded that with increasing $\gamma_{\phi}$, more stationary dark excitons are created, which leads to stronger group-velocity renormalization. This is consistent with theoretical explanations \cite{Ying2024, groenhof2019, sokolovskii2023, fowlerwright2026mapping} for the observed sub-group-velocity transport. 

\section{Low-loss excitation transport via guided polariton channels}
\begin{figure}[t]
    \centering
    \includegraphics[width=\columnwidth]{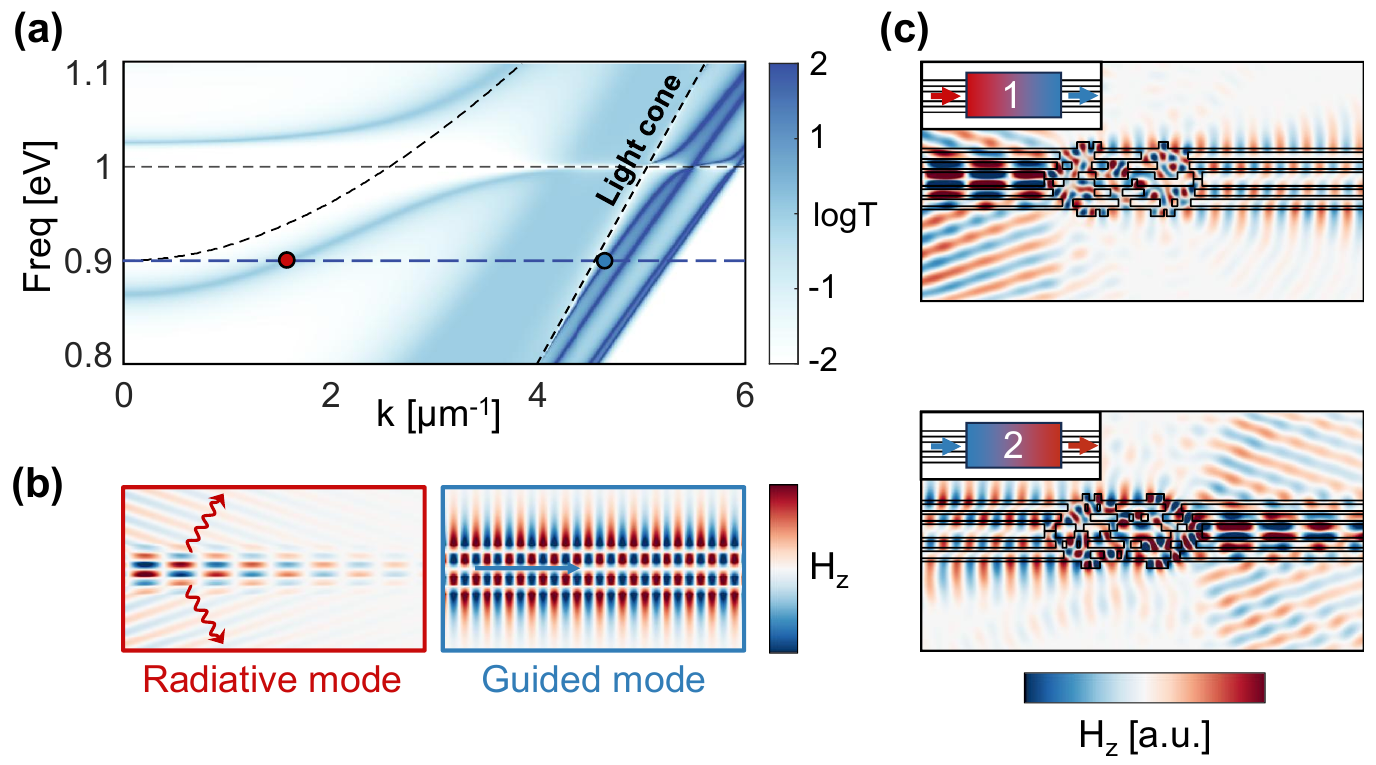}
    \caption{Co-existence of radiative and nonradiative channels inside a DBR cavity. 
    (a) Transmission heatmap of the DBR cavity with molecular layer inside, obtained via TMM. The black dashed lines indicate the molecular transition frequency at $1.0~\mathrm{eV}$, the bare cavity dispersion, and the light line. The blue dashed line marks the operating frequency at $0.9~\mathrm{eV}$, corresponding to the carrier frequency of the input pulse. The red and blue dots identify the cavity polariton mode and the guided polariton mode, respectively. The results are visualized in log scale to reveal the dispersion of the guided modes below the light cone. 
    (b) $H_{z}$ field distribution of the radiative cavity mode and the nonradiative guided mode. 
    (c) Inverse-designed mode converter that transforms the cavity polariton to the guided polariton (top panel), and vice versa (bottom panel). The mode converters' grating structures are indicated by the black lines. 
    }
    \label{fig:fig3}
\end{figure}

In the previous part, we have reproduced known results related to polariton transport: though the LP wave packet travels ballistically, the molecular dephasing transfers energy into stationary dark state populations, leading to sub-group-velocity transport. 
The resulting velocity renormalization increases with higher exciton fraction \cite{xu2023, balasubrahmaniyam2023, jin2023, hong2025excitondelocalization, Ying2024, fowlerwright2026mapping}. 
This limitation appears difficult to overcome, since as long as the excitonic fraction remains nonzero, excitation transport is constrained by the dephasing rate $\gamma_{\phi}$. Such an effect is significant at room temperature and has hindered the construction of polaritonic devices based on organic molecules. 
In this section, we demonstrate that this challenge can be circumvented through structural engineering. 

\subsection{Mode families in DBR cavity}
% We start by pointing out that the above challenge is not fundamental. Instead it comes from the fact that in conventional analysis based on the Tavis-Cummings model, only a single mode family is included (cavity modes described by $\omega_{k} \hat{a}^{\dagger}_{k} \hat{a}_{k}$). 
The trade-off between retaining substantial molecular character and achieving low-loss propagation is not fundamental, but arises when excitation, propagation, and readout are constrained to a single mode family (cavity modes described by $\sum_{k} \omega_{k} \hat{a}^{\dagger}_{k} \hat{a}_{k}$ in the Tavis--Cummings model). 
Within this single mode family, both molecular dephasing and radiative loss impede transport. 
However, such a single-mode-family description does not capture the full electromagnetic environment. 
For example, a standard DBR structure supports not only the radiative cavity modes, but also guided modes below the light line \cite{liu2023, hou2020}. 
The dispersion curves of these modes can be extracted from TMM results. 
As shown in Fig.~\ref{fig:fig3}(a), this system supports two types of modes: the cavity polariton (red dot) as well as the guided polariton (blue dot). 
Analysis of the field profiles of these two modes at $0.9~\mathrm{eV}$, obtained using an eigenmode solver~\cite{sm}, shows that the cavity polariton experiences strong radiative loss as it lies above the light cone $\omega=c_0 k$, where $c_0$ denotes the speed of light in vacuum. In contrast, the guided polariton lies below the light cone and is therefore protected from radiative loss. 
The corresponding field distributions are visualized in Fig.~\ref{fig:fig3}(b), which show how these eigenmodes propagate: without radiative decay, the guided mode decays much slower. 
Furthermore, the guided polariton carries only a small excitonic fraction, and is therefore weakly affected by molecular dephasing, making this channel ideal for low-loss transport. 
Therefore, to take advantage of these low-loss guided modes, a possible strategy is to transfer the excitation into a photon-like guided mode for long-range propagation. 
The cavity polariton is still used for readout since it provides strong optical nonlinearity. 
In this scheme, the guided mode can travel over extended distances with minimal radiative loss before being converted back into a cavity polariton at the final destination for readout.
To realize this strategy, we design two mode converters using topology optimization, as shown in Fig.~\ref{fig:fig3}(c), with details provided in the SM~\cite{sm}. The top panel, denoted converter~1, converts the lossy cavity polariton into a guided mode, whereas converter~2 (the bottom panel), designed independently, performs the reverse conversion from guided to cavity polariton.
These mode converters compensate for the momentum mismatch between the two modes. 
Although for the results presented here, molecules are retained throughout the cavity in Fig.~\ref{fig:fig4}, transport between the two converters is carried mainly
by the photon-like guided mode, so the intermediate region could in principle be molecule-free. 
% Note that since transport relies on the photon-like guided mode, the molecular layer is needed only at the injection and readout ends. The region between the two converters could be left molecule-free. 

\subsection{Long-range transport over millimeter scales}
\begin{figure}[t]
    \centering
    \includegraphics[width=\columnwidth]{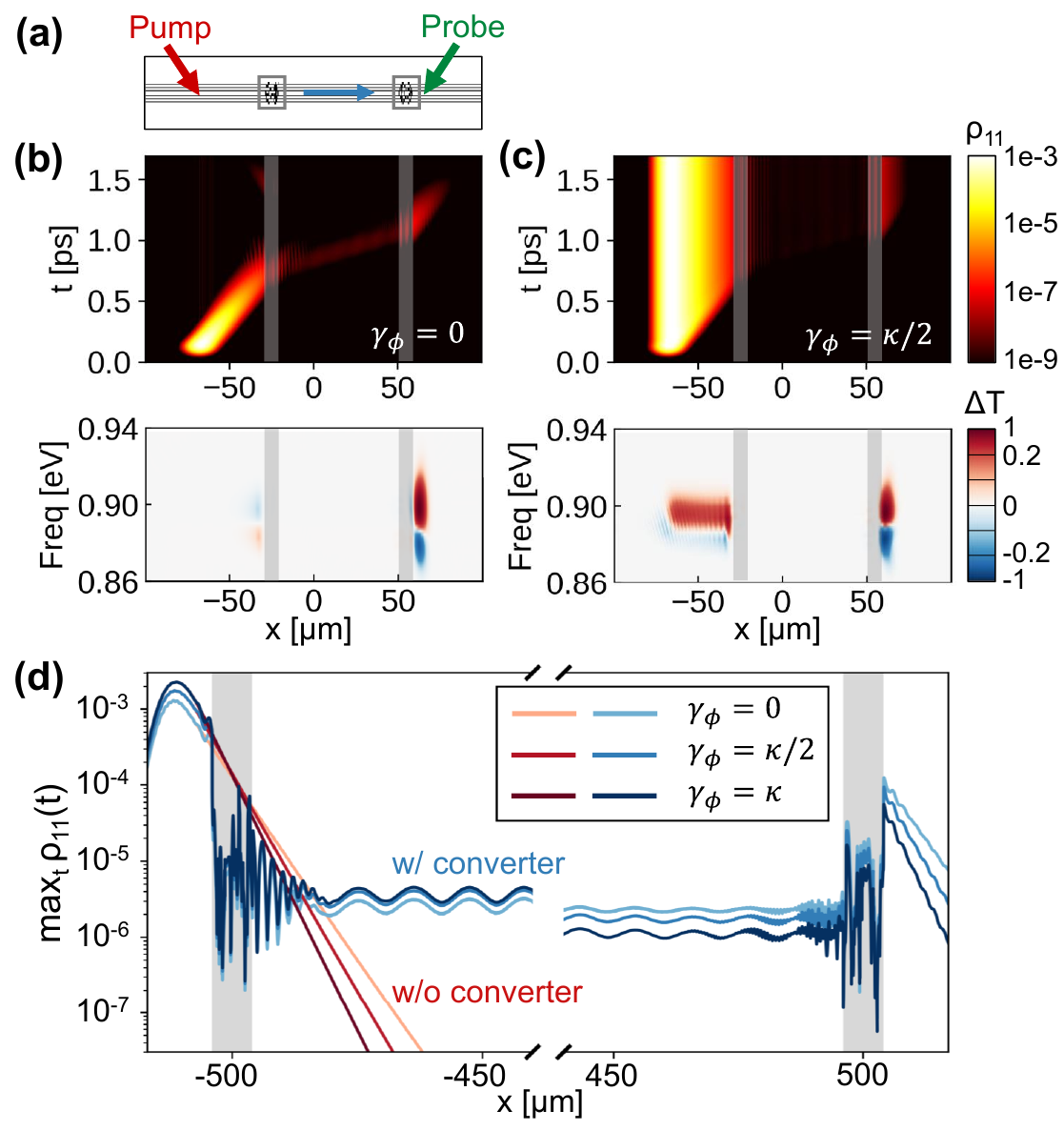}
    \caption{Leveraging the guided mode in a DBR cavity for long-range excitation transport. 
    (a) Simulation setup. The cavity polariton excited by the pump is first converted into a guided polariton by the first converter, propagates through the low-loss guided mode, and is then transformed back into a cavity polariton by the second converter.
    (b) Results without dephasing ($\gamma_{\phi}=0$). Top: population heatmap $\rho_{11}(x, t)$. Bottom: the corresponding pump--probe microscopy signal $\Delta T(x,\omega)$, which is separated into two regions by the mode converters; in the region between them, no radiation escapes the structure, so $\Delta T \approx 0$. % The bright blob that reappears after converter 2 indicates that guided mode is helping excitation transport...
    (c) The same two quantities for $\gamma_{\phi}=\kappa/2$, which leads to static dark population. Top: $\rho_{11}(x, t)$; bottom: $\Delta T(x,\omega)$, which shows a trailing feature associated with dephasing.
    (d) Maximum received molecular excitation $\max_{t} \rho_{11}(x,t)$ over time. With mode converters (blue curves) the excitation drops after converter 1, then rises again after converter 2. For comparison, without mode converters the excitation decays exponentially (red curves). Here the converter separation is $1~\mathrm{mm}$. 
    }
    \label{fig:fig4}
\end{figure}

To demonstrate how these two converters facilitate excitation transport, we resonantly excite a polariton wave packet using a pump pulse, and examine how the wave packet travels when passing through the two converters. 
The simulation setup is shown schematically in Fig.~\ref{fig:fig4}(a): a pump pulse enters at $x=-68~\mu\text{m}$, while the probe pulse enters at $x=68~\mu\text{m}$, with $k_{p}=-k_{p'}=1.41~\mu\text{m}^{-1}$. 
The pump--probe delay is set to $\tau = 1.04~\text{ps}$, so that the probe pulse overlaps the polariton wave packet at the exit of converter~2. %(differential transmission maps for other delays are provided in the SM~\cite{sm}). 
The two converters are separated by $75~\mu\text{m}$. 
We start by turning off dephasing ($\gamma_{\phi}=0$). The results are shown in Fig.~\ref{fig:fig4}(b). 
The top panel shows the heatmap of population $\rho_{11}(x, t)$. 
After hitting converter~1, $\rho_{11}$ becomes much smaller, since the energy has been mostly transformed into guided mode, which only couples weakly to the molecules. 
After traveling $75\,\mu\text{m}$ and hitting the second converter, the $\rho_{11}$ reappears as a bright spot, since the energy inside the guided mode has been transformed back to molecular excitation.
Corresponding coherence dynamics for $|\rho_{10}|$ are provided in the SM~\cite{sm}. 
The bottom panel shows the corresponding differential transmission signal $\Delta T(x, \omega)$ obtained through pump-probe microscopy, similar to Fig.~\ref{fig:fig2}(d). 
The spectral features reveal Rabi contraction as expected. In between the two converters, $\Delta T\approx 0$ since the probe pulse does not leak out when traveling as a guided mode. 
Once the probe travels to the left of converter~1, $\Delta T$ becomes nonzero again, producing the two-blob pattern.
Field snapshots illustrating the radiative and guided propagation pathways are provided in the SM~\cite{sm}. 
The mechanism works well even when $\gamma_{\phi}\neq 0$, since the guided mode is only weakly affected by the molecular dephasing. The results are shown in Fig.~\ref{fig:fig4}(c). 
Notably, the $\Delta T$ pattern shows a trailing feature on the left side of converter 1, since in this region $\rho_{11}>0$ due to the dark state population excited by the pump pulse.
The associated bright and dark state population dynamics are analyzed in the SM~\cite{sm}.

Finally, to demonstrate an ultralong-range excitation transport, we increase the distance between the two converters to $1~\text{mm}$. To demonstrate how the guided mode enables efficient excitation transport, Fig.~\ref{fig:fig4}(d) shows the maximum received excitation, $\max_{t}\rho_{11}(x,t)$, as a function of position $x$. Without mode converters, the excitation decays exponentially with distance, as shown by the red curves. In contrast, when the converters are included, the excitation is restored after converter 2, demonstrating the recovery of molecular excitation following long-range guided-mode propagation.
% A similar effect is also observed in the weak-coupling regime, as explained in the SM~\cite{sm}. 
By exploiting multiple transport channels supported by the DBR structure, we anticipate that excitation transport can be extended beyond the millimeter length scale, more than an order of magnitude longer than state-of-the-art exciton-polariton transport distances reported to date \cite{hou2020, balasubrahmaniyam2023, rozenman2018, pandya2021,liu2023,berghuis2022controlling, liu2023long, Xie20252dmaterial, xu2023, dang2024long, wurdack2021motional}. 
This result shows that the constraints inherent to single-mode-family polariton transport are not fundamental, but can be circumvented by assigning complementary roles to the different mode families naturally hosted by the same structure, thereby enabling injection, long-range transport, and readout within a single planar device. 
% This type of photonic engineering is enabled by the full-wave simulation framework, which is general and naturally accounts for all photonic modes in realistic electromagnetic environments. 

\section{Discussion and conclusions}
% Paragraph 1: summary / takeaway
The formation of exciton-polaritons in the strong-coupling regime enables enhanced excitation transport between molecules. 
While such transport suffers from dephasing, we have shown that this difficulty is not solely a material limitation, but stems from the single mode family assumption, which can be circumvented in realistic structures. 
By considering multiple dispersion branches supported by a standard DBR cavity, energy transport can instead be carried by low-loss, photon-like modes. 
To verify the above idea, we have developed a perturbative Maxwell–Liouville framework for ultrafast nonlinear microscopy. 
By accounting for the feedback of the molecular polarization onto the electromagnetic field in a self-consistent manner, our framework naturally handles multiple mode families and predicts spectroscopic observables in realistic photonic structures. 

% Paragraph 2: Experimental testability
With this simulation framework, we have designed grating structures that serve as mode converters, transforming between cavity polaritons and guided polaritons. 
The proposed scheme should be experimentally achievable: DBR cavities are standard platforms for polariton experiments, and the mode converters require grating structures that can be realized with existing nanofabrication techniques~\cite{phan2019high, piggott2015inverse}. 
A direct experimental signature of the mechanism is the spatial profile of the pump–probe differential transmission: $\Delta T$ should vanish between the two converters, where excitation propagates as a nonradiative guided mode, and reappear beyond the second converter, signaling the recovery of molecular excitation. 
Our simulations demonstrate excitation transport beyond $1~\mathrm{mm}$ even in the presence of molecular dephasing. 

% Paragraph 3: outlook
The formalism offers potential for generalization via multiple avenues. 
First, the FDTD treatment is not restricted to DBR cavities. The same principle can be generalized to more complex photonic structures, such as photonic crystals \cite{ardizzone2022polariton, Xie20252dmaterial} and plasmonic nanoparticles \cite{berghuis2022controlling, jin2023, reitz2026quantum, gonzalez2016uncoupled}, which support even richer mode families. 
Second, while we include pure dephasing by introducing $\gamma_{\phi}$ phenomenologically, a more accurate description of vibronic coupling and non-Markovian dynamics may be essential to understand a richer class of polariton phenomena \cite{fowler2022efficient}.
Furthermore, while we have focused on pump-probe microscopy involving two pulses, the perturbative framework itself generalizes naturally to multidimensional coherent spectroscopy, where different nonlinear pathways can be disentangled \cite{cundiff2013optical, reitz2026multidimensional}. 
The multimodal design principle demonstrated here opens a route toward on-chip polaritonic circuit elements \cite{liu2023, adak2025directionalflow}, as well as reconfigurable polaritonic devices \cite{xiang2019manipulating, liran2025subnanosecond, anton2012dynamics}.
% Incorporating realistic disorder would also be a natural next step since the framework can help us understand how tailoring photonic environment can alleviate the effect of disorder. 

\section{Acknowledgments}
 
P.~F.-W., J.~Y.~Z.~and M.~R.~were primarily supported by the Air Force Office of Scientific Research (AFOSR) through the Multi-University Research Initiative (MURI) program no.~FA9550-22-1-0317. 
Q.~Z. and Z.~Y. were supported by National Science Foundation Grant No. 2016136 for the QLCI center Hybrid Quantum Architectures and Networks. 
Q.~Z. would also like to thank Dr. Ming Zhou for helpful discussions. 

\section{Data availability}
All code used in this study will be made publicly available in a public repository upon acceptance. 
The numerical data supporting the findings of this study are available from the corresponding author upon reasonable request. 

\bibliography{refs}

%putting the Supplement here

\newpage  % Ensure a page break before switching to one-column layout
\onecolumngrid  % Start one-column grid
\clearpage

% Redefine section numbering to include 'S'
\renewcommand{\thesection}{S\arabic{section}}

% Redefine equation numbering to include 'S'
\renewcommand{\theequation}{S\arabic{equation}}

% Redefine figure numbering to include 'S'
\renewcommand{\thefigure}{S\arabic{figure}}

% Redefine table numbering to include 'S' if needed
\renewcommand{\thetable}{S\arabic{table}}

% Reset counters for the supplement/appendix
\setcounter{section}{0}
\setcounter{equation}{0}
\setcounter{figure}{0}
\setcounter{table}{0}
\makeatletter
\let\old@section\section
\renewcommand{\section}{%
  \@ifstar{\old@section*}{\app@section}%
}
\newcommand{\app@section}[1]{%
  \old@section{#1}%
  \addcontentsline{\apptoctype}{section}{\protect\numberline{\thesection}#1}%
}
\makeatother

% remove "Appendix" word from section titles
\renewcommand{\appendixname}{}
\newpage

% Start writing TOC entries into a separate file for the appendix
\addtocontents{\apptoctype}{\protect\setcounter{tocdepth}{2}}

% Add a symmetrically centered title to the appendix
\begin{center}
\begin{large}
Supplemental Material (SM) for\\[4pt]
\textbf{``Full-wave nonlinear microscopy reveals guided channel for ultrafast polariton transport''}
\end{large}
\end{center}

\section{Derivation of Maxwell--Liouville equations}
\label{sec:maxwell_liouville_derivation}

We consider an ensemble of molecules modeled as two-level systems (TLSs) embedded in an electromagnetic environment with arbitrary spatially-dependent permittivity $\varepsilon_r(\bm r)$.  The electromagnetic fields are treated classically, while the molecular degrees of freedom are described quantum mechanically by density matrices $\rho_j$ for molecules located at positions $\bm r_j$. 
Within the coarse-graining, each grid point contains many molecules whose states are assumed to be identical and can therefore be represented by a single density matrix.
This provides an efficient
way of computing the spatially-resolved dynamics with large numbers of molecules. The coupled Maxwell--Liouville equations used in the main text follow from Maxwell's equations with the molecular polarization included as a source term, together with the quantum master equation for each molecule.

We start from the curl equations
\begin{equation}
    \begin{aligned}
    \nabla \times \bm E
    &=
    -\mu_0\partial_t \bm H
    -\bm M_{\mathrm{in}},\\
    \nabla \times \bm H
    &=
    \varepsilon_r\varepsilon_0\partial_t \bm E
    +\bm J_{\mathrm{mol}}
    +\bm J_{\mathrm{in}} .
    \end{aligned}
    \label{eq:S_maxwell_general}
\end{equation}
Here $\bm J_{\mathrm{in}}$ and $\bm M_{\mathrm{in}}$ are externally imposed electric and magnetic current sources used to launch the incident optical pulse.  The molecular current density is obtained from the time derivative of the molecular polarization,
\begin{equation}
    \bm J_{\mathrm{mol}}(\bm r,t)
    =
    \partial_t\bm P_{\mathrm{mol}}(\bm r,t),
    \label{eq:S_Jmol_def}
\end{equation}
where
\begin{equation}
    \bm P_{\mathrm{mol}}(\bm r,t)
    =
    \sum_j
    \Tr\!\left[
    \hat{\bm\mu}_j\rho_j(t)
    \right]
    \delta(\bm r-\bm r_j).
    \label{eq:S_Pmol_def}
\end{equation}

For the two-dimensional simulations considered in this work, the system is invariant along the $z$ direction and we use the transverse-magnetic (TM) polarization, for which the nonzero field components are $(E_x,E_y,H_z)$.  Equation~\eqref{eq:S_maxwell_general} is then written componentwise as
\begin{equation}
\begin{aligned}
    \frac{\partial H_z}{\partial t}
    &=
    \frac{1}{\mu_0}
    \left(
    \frac{\partial E_x}{\partial y}
    -
    \frac{\partial E_y}{\partial x}
    -
    M_{\mathrm{in},z}
    \right),\\
    \frac{\partial E_y}{\partial t}
    &=
    \frac{1}{\varepsilon_r\varepsilon_0}
    \left(
    -\frac{\partial H_z}{\partial x}
    -
    J_{\mathrm{mol},y}
    -
    J_{\mathrm{in},y}
    \right),\\
    \frac{\partial E_x}{\partial t}
    &=
    \frac{1}{\varepsilon_r\varepsilon_0}
    \left(
    \frac{\partial H_z}{\partial y}
    -
    J_{\mathrm{mol},x}
    -
    J_{\mathrm{in},x}
    \right).
\end{aligned}
\label{eq:S_maxwell_TM}
\end{equation}
When a single incident pulse with amplitude $\eta_p$ is used, the external sources are written as
$\bm J_{\mathrm{in}}\rightarrow \eta_p\bm J_{\mathrm{in}}$ and
$\bm M_{\mathrm{in}}\rightarrow \eta_p\bm M_{\mathrm{in}}$,
as in the main text.

The Hamiltonian of molecule $j$ is
\begin{equation}
    H_j(t)
    =
    H_0^j
    +
    H_{\mathrm{int}}^j(t),
    \label{eq:S_mol_hamiltonian}
\end{equation}
with the free and interaction terms
\begin{equation}
    H_0^j
    =
    \frac{\hbar\omega_0}{2}\hat\sigma_j^z,
    \qquad
    H_{\mathrm{int}}^j(t)
    =
    -
    \hat{\bm\mu}_j\cdot
    \bm E(\bm r_j,t).
    \label{eq:S_H0_Hint}
\end{equation}
The molecular dipole operator is
\begin{equation}
    \hat{\bm\mu}_j
    =
    \bm\mu
    \left(
    \hat\sigma_j^-
    +
    \hat\sigma_j^+
    \right),
    \label{eq:S_dipole_operator}
\end{equation}
where the dipole moment $\bm\mu$ is taken to be real.  The density matrix evolves according to the quantum master equation
\begin{equation}
    \frac{d\rho_j}{dt}
    =
    -\frac{i}{\hbar}
    \left[
    H_j(t),\rho_j
    \right]
    +
    \mathcal{D}[\rho_j].
    \label{eq:S_master_general}
\end{equation}
Equations~\eqref{eq:S_maxwell_general} and \eqref{eq:S_master_general}, together with Eq.~\eqref{eq:S_Jmol_def}, constitute the self-consistent Maxwell--Liouville system.

For completeness, we also give the explicit equations for the two-level density-matrix elements.  In the basis $\{\ket{0},\ket{1}\}$, with transition frequency $\omega_0$ and real transition dipole moment $\bm\mu$, the interaction Hamiltonian may be written as
\begin{equation}
    H_{\mathrm{int}}^j(t)
    =
    -
    \bm\mu\cdot\bm E(\bm r_j,t)
    \left(
    \ket{0}_j\bra{1}
    +
    \ket{1}_j\bra{0}
    \right).
    \label{eq:S_interaction_matrix}
\end{equation}
The dissipator $\mathcal{D}[\rho_j]$ is taken to include pure dephasing at rate $\gamma_\phi$. This gives
\begin{equation}
\begin{aligned}
    \frac{d\rho_{11,j}}{dt}
    &=
    \frac{2\,\bm\mu\cdot\bm E(\bm r_j,t)}{\hbar}
    \operatorname{Im}
    \left[
    \rho_{10,j}
    \right],
    \\
    \frac{d\rho_{10,j}}{dt}
    &=
    \left(
    -i\omega_0
    -
    \gamma_\phi
    \right)
    \rho_{10,j}
    +
    \frac{i\,\bm\mu\cdot\bm E(\bm r_j,t)}{\hbar}
    \left(
    \rho_{00,j}
    -
    \rho_{11,j}
    \right),
\end{aligned}
\label{eq:S_density_matrix_elements}
\end{equation}
% with $\rho_{00,j}+\rho_{11,j}=1$ for the full, unexpanded density matrix.  

Finally, the coupling between the molecular and electromagnetic equations is closed through the current source
\begin{equation}
    \bm J_{\mathrm{mol}}(\bm r,t)
    =
    \sum_j
    \partial_t
    \left[
    2\,\bm\mu\,
    \operatorname{Re}
    \rho_{10,j}(t)
    \right]
    \delta(\bm r-\bm r_j),
    \label{eq:S_Jmol_rho10}
\end{equation}
where Eq.~\eqref{eq:S_dipole_operator} has been used.  This source term describes the back-action of the molecular polarization on the electromagnetic field.

\newpage
\section{Details of numerical simulation}
\label{sec:fdtd_scheme}

Here we summarize the time-domain numerical scheme used to evolve the Maxwell--Liouville equations derived in Sec.~\ref{sec:maxwell_liouville_derivation}. 
This section focuses on the Maxwell--Liouville equations used for the pump-only transport simulations.
The perturbative expansion used for pump--probe spectroscopy is described separately in Sec.~\ref{sec:perturbation_expansion}.
% The code uses normalized units internally, with $\varepsilon_0=\mu_0=\hbar=c_0=1$; we keep the physical constants implicit in the material coefficients below. 

% \subsection{Yee grid and time staggering}
The electromagnetic fields are discretized on a uniform Yee-type Cartesian grid. 
While our code supports three-dimensional simulation with all six field components $(E_x,E_y,E_z,H_x,H_y,H_z)$, all simulations displayed in the paper are two-dimensional, thus only $(E_x,E_y,H_z)$ components are nonzero. 
Following the usual leapfrog convention \cite{taflove2005computational}, at the beginning of the $n$-th time step, we start with the electric field $\bm E^{n-\frac{1}{2}}$ at time $(n-\frac{1}{2})\Delta t$, and the magnetic field $\bm H^{n}$ at time $n\Delta t$. 
For all grid points filled with molecules, the corresponding density matrix variables, namely $\rho_{11}^n$ and $\rho_{10}^n$ are also recorded. 
For the transport simulations, the spatial resolution is $\Delta x=17~\mathrm{nm}$. 
The time step is $\Delta t=0.56\,\Delta x/c_0$ so that the Courant–Friedrichs–Lewy condition can be satisfied \cite{taflove2005computational}. 
The simulated time window for the pump-only transport data in Fig.~\ref{fig:fig2} is $0.5$~ps. %corresponding to approximately $1.57\times 10^4$ FDTD steps. 
We impose perfectly matched layers (PMLs) in both $x$ and $y$ directions. 

\begin{figure}[h]
    \centering
    \includegraphics[width=0.35\columnwidth]{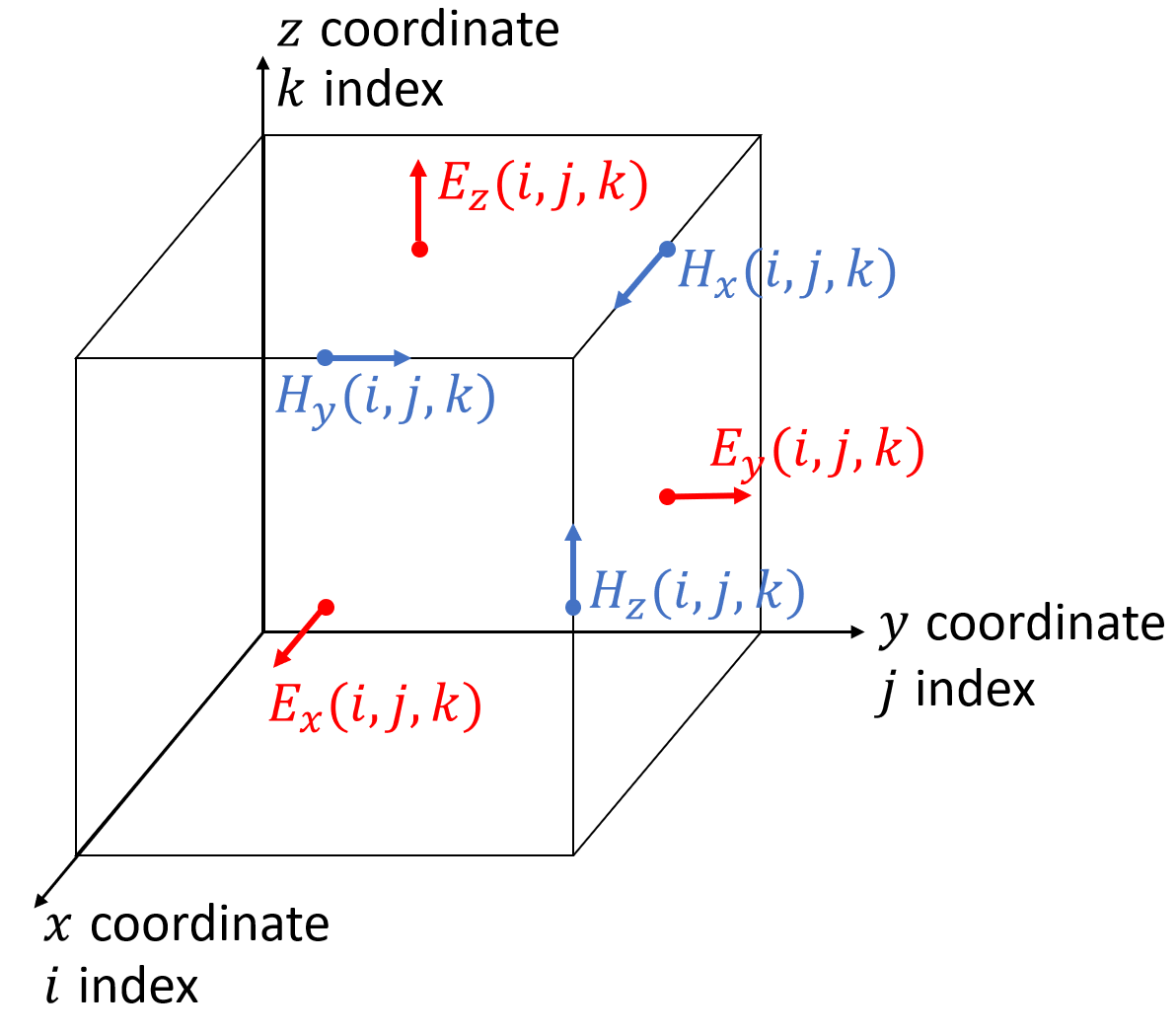}
    \caption{The Yee lattice used in the 3D FDTD simulations. Here only the $(i, j, k)$ grid point is plotted.}
    \label{fig:Yee}
\end{figure}

\subsection{Updating the electromagnetic fields}

The finite-difference update follows the same componentwise form as a standard Yee scheme. 
The current sources $J_x$, $J_y$, and $M_z$ contain both the externally injected sources and the molecular polarization current. 

The electric field is advanced from $n-\frac{1}{2}$ to $n+\frac{1}{2}$:
\begin{equation}
\begin{aligned}
E_x^{n+\frac{1}{2}}(i,j,k)
&=
E_x^{n-\frac{1}{2}}(i,j,k) +
\frac{\Delta t}{\epsilon_{r}(i,j,k)\epsilon_{0}}
\left[
\frac{
H_z^{n}(i,j,k)-H_z^{n}(i,j-1,k)
}{\Delta x}
-
J_x^{n}(i,j,k)
\right],
\end{aligned}
\label{eq:S_fdtd_Ex_update}
\end{equation}

\begin{equation}
\begin{aligned}
E_y^{n+\frac{1}{2}}(i,j,k)
&=E_y^{n-\frac{1}{2}}(i,j,k)
+
\frac{\Delta t}{\epsilon_{r}(i,j,k)\epsilon_{0}}
\left[
-
\frac{
H_z^{n}(i,j,k)-H_z^{n}(i-1,j,k)
}{\Delta x}
-
J_y^{n}(i,j,k)
\right].
\end{aligned}
\label{eq:S_fdtd_Ey_update}
\end{equation}
The magnetic field is then advanced from $n$ to $n+1$ using the updated electric field:
\begin{equation}
\begin{aligned}
H_z^{n+1}(i,j,k)
&=H_z^{n}(i,j,k) +
\frac{\Delta t}{\mu_{0}}
\left[
\frac{
E_x^{n+\frac{1}{2}}(i,j+1,k)
-
E_x^{n+\frac{1}{2}}(i,j,k)
}{\Delta x}
\right.
\\
&\qquad\left.
-
\frac{
E_y^{n+\frac{1}{2}}(i+1,j,k)
-
E_y^{n+\frac{1}{2}}(i,j,k)
}{\Delta x}
-
M_z^{n+\frac{1}{2}}(i,j,k)
\right].
\end{aligned}
\label{eq:S_fdtd_Hz_update}
\end{equation}
% The full CUDA implementation uses the analogous six-component stencil for $(E_x,E_y,E_z,H_x,H_y,H_z)$; the three equations above are the active subset for the simulations in the main text.
Note that the above expressions are simplified compared to the most general form \cite{taflove2005computational} since we have assumed that the DBR cavity consists of lossless, non-magnetic dielectric materials. 

\subsection{TLS update and molecular current}

At each molecular grid cell, the matter degree of freedom is represented by the two density matrix elements $\rho_{11}$ and $\rho_{10}$. 
For any given location $\bm{r}_{j}$, the time-evolution of the density matrix follows
\begin{equation}
\begin{aligned}
    \frac{d\rho_{11,j}}{dt}
    &=
    \frac{2\,\bm\mu\cdot\bm E(\bm r_j,t)}{\hbar}
    \operatorname{Im}
    \left[
    \rho_{10,j}
    \right],
    \\
    \frac{d\rho_{10,j}}{dt}
    &=
    \left(
    -i\omega_0
    -
    \gamma_\phi
    \right)
    \rho_{10,j}
    +
    \frac{i\,\bm\mu\cdot\bm E(\bm r_j,t)}{\hbar}
    \left(
    \rho_{00,j}
    -
    \rho_{11,j}
    \right),
\end{aligned}
% \label{eq:S_density_matrix_elements}
\end{equation}
where $\bm{E}(\bm{r}_{j})$ denotes the local electric field, which serves as a driving term. 
The TLS equations are integrated with a 4th-order Runge--Kutta method. 
Specifically, for a state vector $\bm y=(\rho_{11},\rho_{10})$, we denote the right-hand side as $\bm f(\bm y,\bm E)$, then define
\begin{equation}
\begin{aligned}
\bm k_1 = \bm f(\bm y^n,\bm E),\quad
\bm k_2 = \bm f(\bm y^n+\tfrac{\Delta t}{2}\bm k_1,\bm E),\quad
\bm k_3 = \bm f(\bm y^n+\tfrac{\Delta t}{2}\bm k_2,\bm E),\quad
\bm k_4 = \bm f(\bm y^n+\Delta t\,\bm k_3,\bm E),
\end{aligned}
\end{equation}
which leads to the update rule
\begin{equation}
    \bm y^{n+1}
    =
    \bm y^n
    +
    \frac{\Delta t}{6}
    \left(
    \bm k_1+2\bm k_2+2\bm k_3+\bm k_4
    \right).
\end{equation}
Within a single Runge--Kutta call, the electric field $\bm{E}$ is held fixed. 
The induced molecular polarization on the grid can be calculated as
\begin{equation}
    \bm P_{\mathrm{mol}}
    =
    2\bm \mu\,\operatorname{Re}\rho_{10} \cdot \frac{1}{(\Delta x)^{2}},
    \label{eq:S_fdtd_tls_pol}
\end{equation}
where the factor $1/(\Delta x)^2$ is introduced as a discretized version of the Dirac-$\delta$ function. The spatial resolution is fixed as $\Delta x=17~\mathrm{nm}$. 
The current source that couples back into Maxwell's equations can thus be calculated based on the following finite difference equation: 
\begin{equation}
    \bm J_{\mathrm{mol}}^n
    =
    \frac{
    \bm P_{\mathrm{mol}}^{n+1}
    -
    \bm P_{\mathrm{mol}}^{n-1}
    }{2\Delta t}.
    \label{eq:S_fdtd_tls_current}
\end{equation}
% Only grid cells marked by the molecular mask contribute to $\bm J_{\mathrm{TLS}}$.

\subsection{One Maxwell--Liouville time step}

At the $n$-th time step, suppose that $\bm E^{n-\frac{1}{2}}$, $\bm H^n$, and the TLS state $\rho^n$ are known. 
One full update consists of the following steps: 
\begin{itemize}
    \item Set all current sources to zero. 
    \item Predict the TLS state from $\rho^n$ to $\rho_{\mathrm{pred}}^{n+1}$ using $\bm E^{n-\frac{1}{2}}$.
    \item Convert the predicted coherence to $\bm P_{\mathrm{pred}}^{n+1}$, then calculate the corresponding current source $\bm J_{\mathrm{TLS}}^n$ using Eq.~\eqref{eq:S_fdtd_tls_current}. 
    \item Add the external optical source currents $\bm J_{\mathrm{in}}^n$ and $\bm M_{\mathrm{in}}^{n+\frac{1}{2}}$ to the same current source arrays.
    \item Update the electric field from $\bm E^{n-\frac{1}{2}}$ to $\bm E^{n+\frac{1}{2}}$ using Eqs.~\eqref{eq:S_fdtd_Ex_update} and \eqref{eq:S_fdtd_Ey_update}.
    \item Update the magnetic field from $\bm H^n$ to $\bm H^{n+1}$ using Eq.~\eqref{eq:S_fdtd_Hz_update}.
    \item Correct the TLS state by evolving $\rho^n$ to $\rho^{n+1}$ using the newly updated electric field $\bm E^{n+\frac{1}{2}}$.
\end{itemize}
Note that here we apply a predictor–corrector scheme when updating $\rho$, which ensures numerical stability. 
We implement the above scheme using CUDA C so that it can run in parallel on GPU. 

% (CUDA file-list and planned-figure scaffolding removed)

\newpage
\section{Details of the DBR cavity used}
\label{sec:dbr_cavity}

This section specifies the DBR cavity and molecular layer used for the transport simulations in Fig.~\ref{fig:fig2} and the mode-family analysis in Fig.~\ref{fig:fig3}. 
% In the FDTD implementation the DBR stack is normal to the code coordinate $x$ and the molecular sheet lies along $y$; in the main text and in Secs.~\ref{sec:waveguide_mode}--\ref{sec:inverse_design} propagation is along $x$ and the stack is transverse.  The two differ only by this relabeling. 
The cavity consists of two symmetric DBR mirrors. 
Each DBR is a quarter-wave stack of alternating vacuum ($n_1=1$) and dielectric ($n_2=2$) layers. 
The central frequency is designed as $\omega_c = 0.9$~eV ($\lambda_c=1377.6$~nm), which gives quarter-wave thicknesses $t_1=\lambda_c/(4 n_1)=344.4~$nm and $t_2=\lambda_c/(4 n_2)=172.2~$nm. 
We use $N_{\mathrm{DBR}}=3$ periods on each side of the cavity. 
The cavity shows a photon loss rate of $\kappa\approx 0.0102$~eV, extracted from the transmission spectrum at normal incidence. 

The molecular layer is represented by a single layer of two-level systems placed at the cavity center.
The transition frequency is fixed as $\omega_0=1.0$~eV ($\lambda_0=1239.8$~nm). 
While each grid point contains many molecules, the states of all molecules inside a grid point are assumed to be identical, leading to a spatial coarse-graining. 
The dipole moments of molecules are aligned along the $x$-axis. 

% We vary the pure dephasing rate between $\gamma_\phi=0$ and $\gamma_\phi=0.005\,\omega_0\simeq\kappa/2$. 

% strong-coupling regime. 
% permittivity of molecular layer

\begin{figure*}[h]
    \centering
    \includegraphics[width=1.0\textwidth]{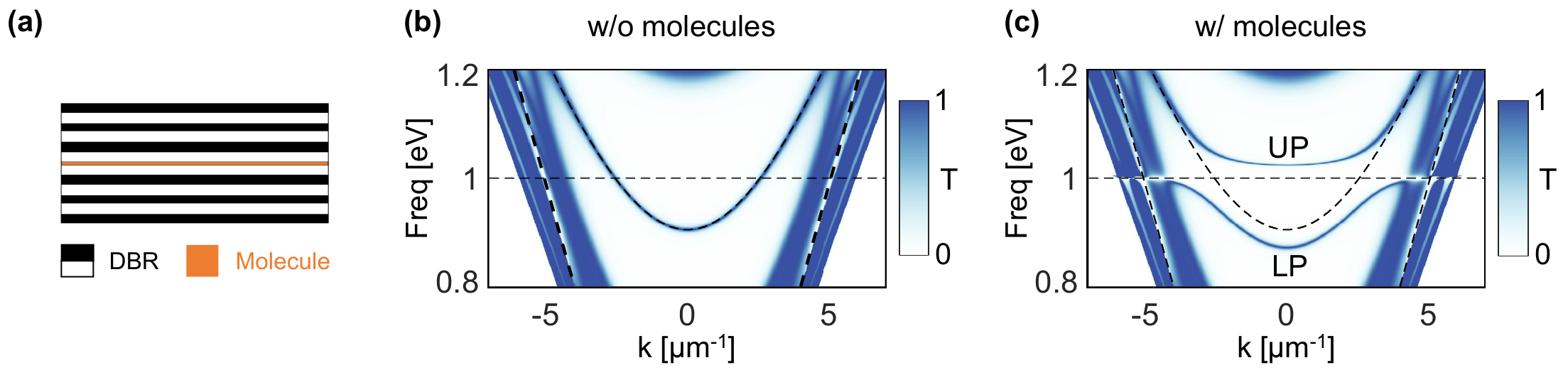}
    \caption{TMM results. 
    (a) Schematic illustration of the DBR cavity. 
    (b) Transmission spectrum of bare cavity. 
    (c) Transmission spectrum of cavity, with molecular layer inside. The heatmap reveals the upper polariton (UP) and the lower polariton (LP). 
    Both spectra have been calculated using TMM. }
    \label{fig_SI:DBR_TMM_result}
\end{figure*}

The dispersion relation of the modes supported by the DBR cavity can be obtained using the transfer-matrix method (TMM). 
Specifically, we vary the in-plane wave vector of an incident plane wave and measure the transmission. 
The result for the bare cavity is shown in Fig.~\ref{fig_SI:DBR_TMM_result}(b). 
With the molecular layer inside, the transmission spectrum reveals the upper and lower polariton branches, as shown in Fig.~\ref{fig_SI:DBR_TMM_result}(c). 

\newpage
\section{Derivation of perturbation expansion: pump--probe}
\label{sec:perturbation_expansion}

Here we derive the perturbative Maxwell--Liouville hierarchy used for the computation of the pump--probe response.  We consider two incident fields, a pump $p$ and a probe $p'$, with amplitudes $\eta_p$ and $\eta_{p'}$, respectively. The self-consistent Maxwell's equations with the two input pulses are
\begin{equation}
    \begin{aligned}
    \nabla \times \bm{E}
    &=
    -\mu_{0}\partial_{t}\bm{H}
    -\eta_p \bm{M}_{\mathrm{in}}^{p}
    -\eta_{p'} \bm{M}_{\mathrm{in}}^{p'},\\
    \nabla \times \bm{H}
    &=
    \varepsilon_r\varepsilon_0 \partial_t \bm{E}
    +\bm{J}_{\mathrm{mol}}
    +\eta_p \bm{J}_{\mathrm{in}}^{p}
    +\eta_{p'} \bm{J}_{\mathrm{in}}^{p'} .
    \end{aligned}
    \label{eq:S_maxwell_pp}
\end{equation}
The molecular current is generated by the dipole polarization,
\begin{equation}
    \bm J_{\mathrm{mol}}(\bm r,t)
    =
    \sum_j
    \partial_t
    \Tr\!\left[
    \hat{\bm\mu}_j \rho_j(t)
    \right]
    \delta(\bm r-\bm r_j).
    \label{eq:S_mol_current}
\end{equation}
The density matrix of molecule $j$ evolves according to
\begin{equation}
    \frac{d\rho_j}{dt}
    =
    -\frac{i}{\hbar}
    \left[
    H_0^j
    -\hat{\bm\mu}_j\cdot \bm E(\bm r_j,t),
    \rho_j
    \right]
    +\mathcal{D}[\rho_j].
    \label{eq:S_liouville_pp}
\end{equation}

We expand both the electromagnetic fields and the molecular density matrix in powers of the pump and probe amplitudes,
\begin{equation}
\begin{aligned}
    \{\bm E,\bm H\}
    &=
    \sum_{n=0}^{\infty}
    \sum_{n'=0}^{\infty}
    \eta_p^{\,n}\eta_{p'}^{\,n'}
    \{\bm E^{(n)(n')},\bm H^{(n)(n')}\},\\
    \rho_j
    &=
    \sum_{n=0}^{\infty}
    \sum_{n'=0}^{\infty}
    \eta_p^{\,n}\eta_{p'}^{\,n'}
    \rho_j^{(n)(n')}.
\end{aligned}
\label{eq:S_pp_expansion}
\end{equation}
Substitution into Eq.~\eqref{eq:S_maxwell_pp} gives the Maxwell's equations at fixed perturbative order,
\begin{equation}
    \begin{aligned}
    \nabla \times \bm{E}^{(n)(n')}
    &=
    -\mu_{0}\partial_t \bm{H}^{(n)(n')}
    -\delta_{1n}\delta_{0n'}\bm{M}_{\mathrm{in}}^{p}
    -\delta_{0n}\delta_{1n'}\bm{M}_{\mathrm{in}}^{p'},\\
    \nabla \times \bm{H}^{(n)(n')}
    &=
    \varepsilon_r\varepsilon_0 \partial_t \bm{E}^{(n)(n')}
    +\bm{J}_{\mathrm{mol}}^{(n)(n')}
    +\delta_{1n}\delta_{0n'}\bm{J}_{\mathrm{in}}^{p}
    +\delta_{0n}\delta_{1n'}\bm{J}_{\mathrm{in}}^{p'} .
    \end{aligned}
    \label{eq:S_maxwell_order}
\end{equation}
Here
\begin{equation}
    \bm J_{\mathrm{mol}}^{(n)(n')}(\bm r,t)
    =
    \sum_j
    \partial_t
    \Tr\!\left[
    \hat{\bm\mu}_j \rho_j^{(n)(n')}(t)
    \right]
    \delta(\bm r-\bm r_j).
    \label{eq:S_mol_current_order}
\end{equation}
Thus, the incident pump and probe appear only at orders $(1)(0)$ and $(0)(1)$, while all higher-order electromagnetic fields are generated by the molecular polarization. Since for a TLS, at even orders only population and no polarization is created, all even-order electromagnetic fields are vanishing.

The density-matrix equation at order $(n)(n')$ is
\begin{equation}
\begin{aligned}
    \frac{d\rho_j^{(n)(n')}}{dt}
    &=
    -\frac{i}{\hbar}
    [H_0^j,\rho_j^{(n)(n')}]
    +\mathcal{D}[\rho_j^{(n)(n')}]\\
    &\quad
    +\frac{i}{\hbar}
    \sum_{m=0}^{n}
    \sum_{m'=0}^{n'}
    \left[
    \hat{\bm\mu}_j\cdot
    \bm E^{(n-m)(n'-m')}(\bm r_j,t),
    \rho_j^{(m)(m')}
    \right].
\end{aligned}
\label{eq:S_rho_order}
\end{equation}
This equation makes explicit the hierarchical structure of the perturbation expansion: the density matrix at order $(n)(n')$ is driven by products of fields and density matrices whose total order is $(n)(n')$.

For the TLS used in the simulations, we write the density-matrix elements as $\rho_{\alpha\beta,j}^{(n)(n')}$, with $\alpha,\beta\in\{0,1\}$.  Including pure dephasing at rate $\gamma_\phi$, Eq.~\eqref{eq:S_rho_order} becomes
\begin{equation}
\begin{aligned}
    \frac{d\rho_{11,j}^{(n)(n')}}{dt}
    &=
    \sum_{m=0}^{n}
    \sum_{m'=0}^{n'}
    \frac{2\,\bm\mu\cdot
    \bm E^{(n-m)(n'-m')}(\bm r_j,t)}{\hbar}
    \operatorname{Im}
    \left[
    \rho_{10,j}^{(m)(m')}
    \right],
    \\
    \frac{d\rho_{10,j}^{(n)(n')}}{dt}
    &=
    \left(
    -i\omega_0-\gamma_\phi
    \right)
    \rho_{10,j}^{(n)(n')}
    \\
    &\quad
    +
    \sum_{m=0}^{n}
    \sum_{m'=0}^{n'}
    \frac{i\,\bm\mu\cdot
    \bm E^{(n-m)(n'-m')}(\bm r_j,t)}{\hbar}
    \left[
    \rho_{00,j}^{(m)(m')}
    -
    \rho_{11,j}^{(m)(m')}
    \right].
\end{aligned}
\label{eq:S_bloch_order}
\end{equation}

\begin{figure*}[t]
    \centering
    \includegraphics[width=0.9\textwidth]{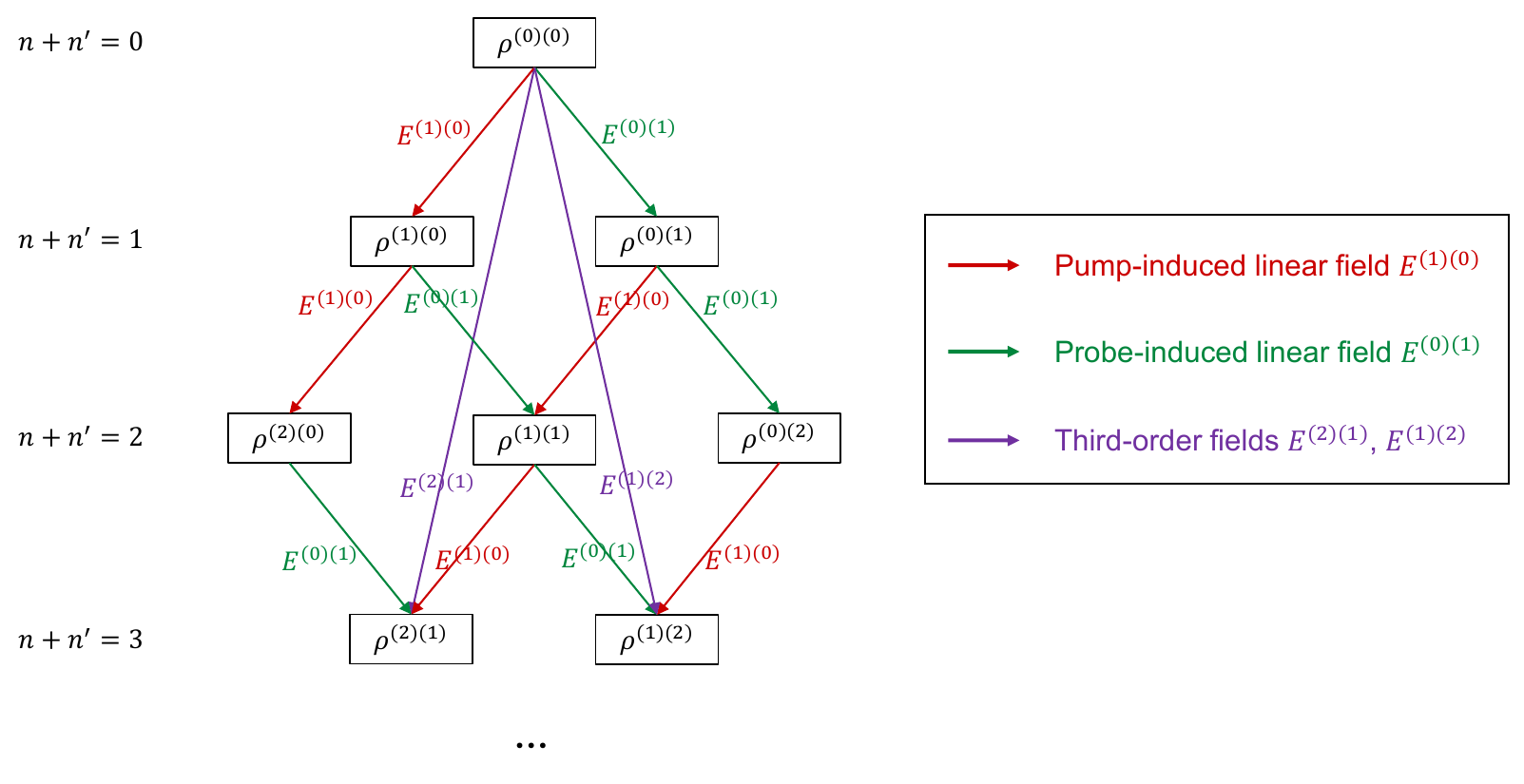}
    \caption{``Subway map'' for full-wave nonlinear microscopy showing the perturbative hierarchy of higher-order density matrices constructed from lower-order density matrices via linear and nonlinear electromagnetic fields. Different pathways are marked with different colors. Note that only nonzero terms are visualized, while vanishing even-order fields (such as $E^{(2)(0)},E^{(1)(1)},\hdots$) are omitted. Higher-order terms not used in this paper are also neglected.}
    \label{fig_SI:perturbation_hierarchy}
\end{figure*}

We now list the equations involved in this work for specific $(n)(n')$ orders. The construction of higher-order density matrices from lower-order density matrices
is illustrated in Fig.~\ref{fig_SI:perturbation_hierarchy}. 
We start from the zeroth-order quantities, which are fixed by the initial state of the system. In particular,
\begin{equation}
    \rho_{\alpha\beta,j}^{(0)(0)}(0)
    =
    \rho_{\alpha\beta,j}(0),
    \qquad
    \bm E^{(0)(0)}(\bm r,0)
    =
    \bm E(\bm r,0),
    \qquad
    \bm H^{(0)(0)}(\bm r,0)
    =
    \bm H(\bm r,0),
\end{equation}
while all higher-order components are initialized to zero.  For the simulations considered here, the molecules are initially in the ground state and no electromagnetic field is present, so that
\begin{equation}
    \rho_{00,j}^{(0)(0)}(0)=1,\qquad
    \rho_{11,j}^{(0)(0)}(0)
    =
    \rho_{10,j}^{(0)(0)}(0)=0,
    \qquad
    \bm E^{(0)(0)}(\bm r,0)
    =
    \bm H^{(0)(0)}(\bm r,0)
    =
    0 .
\end{equation}

The orders relevant to the lowest-order pump--probe signal are obtained sequentially.  The linear probe and pump coherences obey
\begin{equation}
    \frac{d\rho_{10,j}^{(0)(1)}}{dt}
    =
    \left(
    -i\omega_0-\gamma_\phi
    \right)
    \rho_{10,j}^{(0)(1)}
    +
    \frac{i\,\bm\mu\cdot\bm E^{(0)(1)}(\bm r_j,t)}{\hbar},
    \label{eq:S_probe_only}
\end{equation}
and
\begin{equation}
    \frac{d\rho_{10,j}^{(1)(0)}}{dt}
    =
    \left(
    -i\omega_0-\gamma_\phi
    \right)
    \rho_{10,j}^{(1)(0)}
    +
    \frac{i\,\bm\mu\cdot\bm E^{(1)(0)}(\bm r_j,t)}{\hbar}.
    \label{eq:S_pump_only}
\end{equation}
The pump-induced population at order $(2)(0)$ is
\begin{equation}
    \frac{d\rho_{11,j}^{(2)(0)}}{dt}
    =
    \frac{2\,\bm\mu\cdot\bm E^{(1)(0)}(\bm r_j,t)}{\hbar}
    \operatorname{Im}
    \left[
    \rho_{10,j}^{(1)(0)}
    \right].
    \label{eq:S_pump_population}
\end{equation}
Similarly, the mixed pump--probe population at order $(1)(1)$ is
\begin{equation}
\begin{aligned}
    \frac{d\rho_{11,j}^{(1)(1)}}{dt}
    &=
    \frac{2\,\bm\mu\cdot\bm E^{(1)(0)}(\bm r_j,t)}{\hbar}
    \operatorname{Im}
    \left[
    \rho_{10,j}^{(0)(1)}
    \right]\\
    &\quad
    +
    \frac{2\,\bm\mu\cdot\bm E^{(0)(1)}(\bm r_j,t)}{\hbar}
    \operatorname{Im}
    \left[
    \rho_{10,j}^{(1)(0)}
    \right].
\end{aligned}
\label{eq:S_mixed_population}
\end{equation}

The leading pump-induced correction to the probe transmission is the field at order $(2)(1)$.  Its molecular source is determined by
\begin{equation}
\begin{aligned}
    \frac{d\rho_{10,j}^{(2)(1)}}{dt}
    &=
    \left(
    -i\omega_0-\gamma_\phi
    \right)
    \rho_{10,j}^{(2)(1)}
    +
    \frac{i\,\bm\mu\cdot
    \bm E^{(2)(1)}(\bm r_j,t)}{\hbar}
    \\
    &\quad
    +
    \frac{i\,\bm\mu\cdot
    \bm E^{(1)(0)}(\bm r_j,t)}{\hbar}
    \left[
    \rho_{00,j}^{(1)(1)}
    -
    \rho_{11,j}^{(1)(1)}
    \right]
    \\
    &\quad
    +
    \frac{i\,\bm\mu\cdot
    \bm E^{(0)(1)}(\bm r_j,t)}{\hbar}
    \left[
    \rho_{00,j}^{(2)(0)}
    -
    \rho_{11,j}^{(2)(0)}
    \right].
\end{aligned}
\label{eq:S_third_order_probe}
\end{equation}
Eqs.~\eqref{eq:S_maxwell_order} and \eqref{eq:S_third_order_probe} determine $\bm E^{(2)(1)}$ and $\bm H^{(2)(1)}$, which enter the leading pump--probe differential transmission discussed in the main text.  In the FDTD implementation, the required orders are propagated concurrently, with each order receiving its electromagnetic source from $\bm J_{\mathrm{mol}}^{(n)(n')}$ and its nonlinear molecular driving from lower-order fields and density matrices. 

\newpage
\section{Details for calculating $\Delta T$ at different $(n, n')$ order}
\label{sec:T_diff_calc}

The pump--probe signal is defined as the difference between the probe
transmission with and without the pump,
\begin{equation}
    \Delta T_{p'}(\bm r,\omega)
    =
    T^{\mathrm{pump\,on}}_{p'}(\bm r,\omega)
    -
    T^{\mathrm{pump\,off}}_{p'}(\bm r,\omega).
\end{equation}
For the TM fields used here, the transmitted flux through a plane normal to
$y$ is proportional to the $y$ component of the Poynting vector,
\begin{equation}
    S_y(\bm r,\omega)
    =
    \frac{1}{2}\operatorname{Re}
    [\bm E(\bm r,\omega)\times \bm H^*(\bm r,\omega)]_y
    =
    -\frac{1}{2}\operatorname{Re}
    \left[
    E_x(\bm r,\omega)H_z^*(\bm r,\omega)
    \right].
\end{equation}

The probe-direction fields without the pump are, to lowest order,
\begin{equation}
    E_x^{\mathrm{pump\,off}}
    =
    \eta_{p'}E_x^{(0)(1)}, 
    \qquad
    H_z^{\mathrm{pump\,off}}
    =
    \eta_{p'}H_z^{(0)(1)} .
\end{equation}
With the pump present, the leading pump-induced correction to the probe field
appears at order $(2)(1)$:
\begin{equation}
    E_x^{\mathrm{pump\,on}}
    =
    \eta_{p'}E_x^{(0)(1)}
    +
    \eta_p^2\eta_{p'}E_x^{(2)(1)},
    \qquad
    H_z^{\mathrm{pump\,on}}
    =
    \eta_{p'}H_z^{(0)(1)}
    +
    \eta_p^2\eta_{p'}H_z^{(2)(1)} .
\end{equation}

\begin{figure*}[t]
    \centering
    \includegraphics[width=0.7\textwidth]{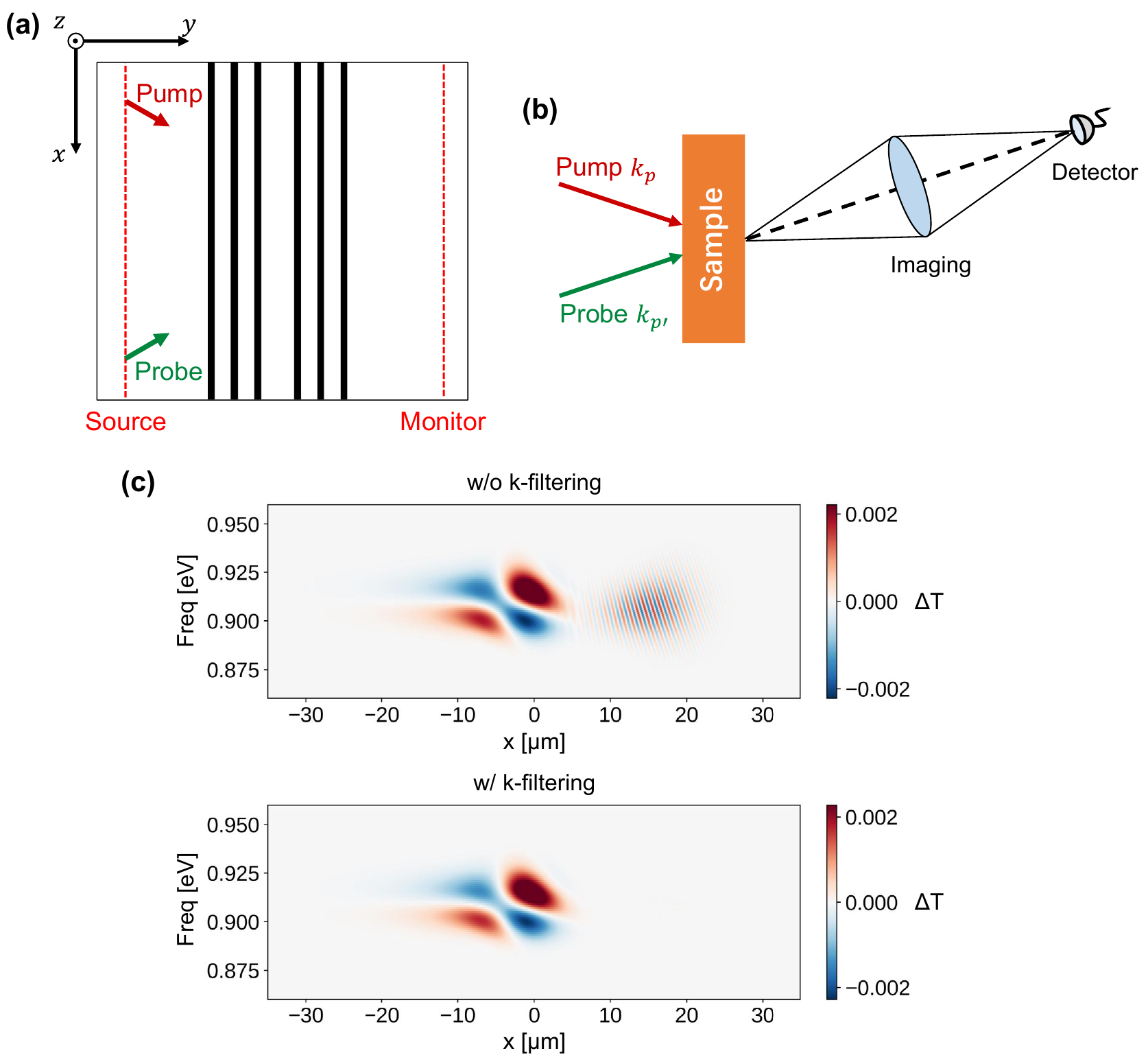}
    \caption{$k$-space filtering before calculating the differential transmission. 
    (a) The simulation setup. The source plane and the monitor plane are marked. 
    (b) Schematic illustration of a realistic experimental setup. An objective lens with finite NA is used to collect the signal, which justifies the $k$-space filtering. 
    (c) Effect of the $k$-space filtering: the fast-oscillating pattern arising from the double-quantum-coherence contribution at $k_x = 3k_p$ is removed, thereby isolating the pump--probe signal along the probe direction $k_{p'}$.
    }
    \label{fig_SI:k_filter}
\end{figure*}

Substituting these fields into the Poynting flux and subtracting the pump-off
contribution gives, after normalization by the incident probe intensity
$\propto \eta_{p'}^2$,
\begin{equation}
\begin{aligned}
    \Delta T(\bm r,\omega)
    &\propto
    -\frac{\eta_p^2}{2}
    \operatorname{Re}
    \Big[
    E_x^{(0)(1)}(\bm r,\omega)
    H_z^{(2)(1)*}(\bm r,\omega)
    \\
    &\qquad\qquad
    +
    E_x^{(2)(1)}(\bm r,\omega)
    H_z^{(0)(1)*}(\bm r,\omega)
    \Big].
\end{aligned}
\end{equation}
Thus, the leading differential transmission comes from the interference between the linear probe field $(0)(1)$ (the local oscillator) and the pump-induced correction to the probe field $(2)(1)$. Higher-order terms, such as $E_x^{(2)(1)}H_z^{(2)(1)*}$, scale as $\eta_p^4$ and are neglected. 
The fields are recorded at the monitor plane, as shown schematically in Fig.~\ref{fig_SI:k_filter}(a). 

% \begin{itemize}
%     \item \mike{explain effect of $k$-space filtering/phase matching and different contributions in momentum space}
% \end{itemize}

The $\Delta T(\bm r,\omega)$ above is not yet the differential transmission signal displayed in the main text. 
In a typical pump--probe microscopy setup, detectors collect only the light emitted into a finite angular range around a given direction.  
This angular range is set by the numerical aperture (NA) of the objective lens, as shown in Fig.~\ref{fig_SI:k_filter}(b). 
% and a fixed emission angle $\theta$ maps one-to-one onto a fixed in-plane momentum $k_x = (\omega/c_0)\sin\theta$. 
We therefore Fourier transform the monitored fields along the propagation direction $x$ and keep only the components co-propagating with the probe. 
% To see why this selection is required, we track the in-plane momentum carried by each order. 
Specifically, in our setup where the pump and probe are launched from opposite sides, the linear probe field at order $(0)(1)$ is centered at $k_x=k_{p'}=-k_p$. 
The third-order field at order $(2)(1)$, generated by two pump interactions and one probe interaction, contains two independent positive-frequency phase-matched components, together with their complex-conjugate counterparts \cite{mukamel1995principles}. 
One component corresponds to the ordinary pump--probe pathway, with in-plane momentum combinations $\pm k_p\mp k_p+k_{p'}=k_{p'}=-k_p$, and is therefore co-propagating with the probe. 
Additional phase components arise from the so-called double-quantum-coherence pathways $2k_p-k_{p'}=3k_p$, which propagate in a different direction and produce the rapidly oscillating spatial pattern shown in Fig.~\ref{fig_SI:k_filter}(c). 
Thus, the $k$-space filter isolates the desired probe-direction signal while removing contributions emitted into other phase-matched directions~\cite{reitz2025nonlinear,fowlerwright2026mapping}. 
% The remaining sidebands are four-wave-mixing signals radiated into different directions; a probe-direction detector does not collect them, and if retained their cross term oscillates rapidly in $x$ and obscures $\Delta T$. 
We implement this selection by introducing a per-frequency Gaussian window in $k_x$. 
For each frequency, the electromagnetic field is first transformed into $k$-space as $\tilde f(k_x,\omega)=\mathcal{F}_x[f(x,\omega)]$, then filtered by a Gaussian kernel
\begin{equation}
    M(k_x) = \exp\!\left[-\frac{1}{2}
    \left(\frac{k_x - k_c}{\Delta k}\right)^{2}\right],
    \qquad
    k_c = k_{p'},\quad \Delta k = 0.6\,k_\parallel,
    \label{eq:S_kfilter}
\end{equation}
and transformed back to real space. 
The same filtering is applied to $E_x^{(0)(1)},H_z^{(0)(1)}$ and $E_x^{(2)(1)},H_z^{(2)(1)}$, before they are used to calculate $\Delta T(x, \omega)$. 
Fig.~\ref{fig_SI:k_filter}(c) illustrates the effect of the $k$-space filtering used to select the probe-direction phase-matched component \cite{mukamel1995principles, Hamm_Zanni_2011}. 
Displayed are the differential transmission results $\Delta T(x, \omega)$ obtained using a similar setup as that used in Fig.~\ref{fig:fig2}(d), only with $\gamma_{\phi}=0$. 
Without the filtering, $\Delta T(x,\omega)$ includes a rapidly oscillating term arising from interference between the double-quantum-coherence component at $k_x=3k_p$ and the linear probe field. 
In a realistic experiment, this off-axis component lies outside the finite collection range of the objective lens.
% Physically, $M(k_x)$ models an angle-resolved detector centered on the probe direction with a finite numerical aperture; for the Fig.~\ref{fig:fig2} parameters ($k_\parallel = \pi/2~\mu\mathrm{m}^{-1}$) at $0.9~\mathrm{eV}$, this corresponds to a central emission angle $\theta_c \approx 20^\circ$ with an acceptance of $\approx\pm 13^\circ$ (one standard deviation).
% The injected in-plane wavevector is $k_\parallel = \pi/2~\mu\mathrm{m}^{-1}$ for the bare-cavity results (Fig.~\ref{fig:fig2}) and $k_\parallel = 1.41~\mu\mathrm{m}^{-1}$ for the mode-converter results (Figs.~\ref{fig:fig3}--\ref{fig:fig4}); the filter center $k_c$ is set to $k_{p'}=-k_\parallel$ accordingly.

\newpage
\section{Extraction of transport velocity from $\Delta T$}
\label{sec:velocity_extraction}

Here we describe how the velocity data points in Fig.~\ref{fig:fig2}(e) are extracted from the simulated differential-transmission maps. 
In the previous Sec.~\ref{sec:T_diff_calc}, we have already explained how $\Delta T(\bm r,\omega)$ can be calculated. 
Note that the $\Delta T(\bm r,\omega)$ profile is only calculated at the output plane, which is located at $y=5.67~\mu$m. 
Thus, for a given pump--probe time delay $\tau$, we simplify the notation as $\Delta T(x, \omega; \tau)$. 
% where $x$ denotes the in-plane coordinate along the output monitor, $\tau$ is the pump--probe delay, $k_p$ is the pump in-plane wave vector, and $\gamma_\phi$ is the molecular pure dephasing rate. 

For each $(k_p,\gamma_\phi)$ case, the simulation is repeated for five delays $\tau \in\{-0.30,-0.15,0,0.15,0.30\}~\mathrm{ps}$. 
% Before the velocity extraction, the post-processing selects the transmitted probe-like component in momentum space, centered at the opposite in-plane wave vector from the pump.  The velocity extraction itself only uses the resulting real-space map $\Delta T(\omega,x)$.
For each delay $\tau$, the differential-transmission map $\Delta T(x, \omega; \tau)$ is first reduced to a one-dimensional spatial envelope by summing over the monitored frequencies:
\begin{equation}
    S(x;\tau)
    =
    \sum_m
    \left|
    \Delta T(x, \omega_m;\tau)
    \right|.
    \label{eq:S_velocity_spatial_signal}
\end{equation}
% The absolute value is taken before summing over frequency so that positive and negative lobes of the differential signal contribute to the same spatial feature rather than canceling each other. 
The peak position can then be extracted by looking for the maximum value:
\begin{equation}
    x_{\mathrm{peak}}(\tau)
    =
    \operatorname*{arg\,max}_{x}
    S(x;\tau).
    \label{eq:S_velocity_peak_position}
\end{equation}
This metric directly tracks the most intense feature in the $\Delta T$ map. 
We also compute a root-mean-square (rms) displacement relative to a fixed reference position $x_0=-25~\mu\mathrm{m}$. 
The rms coordinate can be calculated as
\begin{equation}
    x_{\mathrm{rms}}(\tau)
    =
    \left[
    \frac{
    \sum_j
    S(x_j;\tau)\cdot
    (x_j-x_0)^2
    }{
    \sum_j
    S(x_j;\tau)
    }
    \right]^{1/2}.
    \label{eq:S_velocity_rms_position}
\end{equation}

For each fixed $(k_p,\gamma_\phi)$ and for each position metric $\alpha\in\{\mathrm{peak},\mathrm{rms}\}$, the delay-dependent position is fitted to a straight line,
\begin{equation}
    x_{\alpha}(\tau_i)
    \simeq
    k_{\alpha}\cdot\tau_i
    +
    b_{\alpha}.
    \label{eq:S_velocity_linear_fit}
\end{equation}
The velocity reported in Fig.~\ref{fig:fig2}(e) is $v_{\alpha}=2k_{\alpha}$. 
The factor of two comes from the symmetric pump--probe geometry used in Fig.~\ref{fig:fig2}: pump and probe pulses are launched with opposite in-plane momenta and travel towards each other. 
Repeating the above procedure for $k_p=1.0,1.2,\ldots,2.6~\mu\mathrm{m}^{-1}$ and $\gamma_\phi\in\{0,\kappa/4,\kappa/2\}$ yields the data shown in Fig.~\ref{fig:fig2}(e). % The monitored frequency window is $0.84$--$0.99$~eV with a spacing of $0.001$~eV.

% \newpage
% \section{Diffusion of $k\approx 0$: the effect of mode converter?}
% Note: I commented this out because right now the converters are designed for a given k vector so it does not really make sense to test it around k=0. We can discuss this later if needed. 

\newpage
\section{Waveguide eigenmode solver}
\label{sec:waveguide_mode}

Here we describe how the mode profiles shown in Fig.~\ref{fig:fig3}(b) are calculated. 
The DBR structure is invariant along the propagation direction $x$, while the multilayer stack is arranged along the $y$-axis. 
We solve for the TM eigenmodes that propagate along the $x$-axis, with nonzero field components $(E_x,E_y,H_z)$. %and take the time dependence to be $\exp(-i\omega t)$. 
At any given optical frequency $\omega$, a propagating eigenmode can be described as
\begin{equation}
    H_z(x,y,t)
    =
    \psi(y) \cdot
    \exp(i\beta x-i\omega t),
    % \qquad
    % n_{\mathrm{eff}}
    % =
    % \frac{\beta}{k_0},
    % \qquad
    % k_0=\frac{\omega}{c_0}.
    \label{eq:S_wg_mode_ansatz}
\end{equation}
% With this sign convention, a positive imaginary part of $n_{\mathrm{eff}}$ gives propagation decay proportional to $\exp[-\operatorname{Im}(n_{\mathrm{eff}})k_0x]$. 
where $ \exp(i\beta x)$ and $\psi(y)$ describe the spatial dependence in $x$ and $y$ direction, respectively. 
In frequency domain, Maxwell's equations can be simplified as
\begin{equation}
\frac{\partial^{2}}{\partial x^{2}} H_{z} 
+ \varepsilon_{yy}\frac{\partial}{\partial y} \left( \varepsilon_{xx}^{-1}\frac{\partial H_z}{\partial y} \right)
+ \varepsilon_{yy} k_{0}^{2} H_{z} = 0,
    \label{eq:S_TM_master}
\end{equation}
where $k_{0}=\omega / c_{0}$ denotes the wave vector inside vacuum. 
The anisotropic permittivity of the molecules is calculated based on the following equations:
\begin{equation}
    \begin{aligned}
        \varepsilon_{xx} &= 1 - \frac{2\mu^{2}}{\hbar \varepsilon_{0} (\Delta x)^{2}} \cdot \frac{1}{(\omega - \omega_{0}) + i\gamma_{\phi}}, \\
        \varepsilon_{yy} &= 1, 
    \end{aligned}
\end{equation}
since the dipole moments are all aligned with the $x$-axis in our case. We set $\omega_{0}=1.0~\mathrm{eV}$. The proposed methodology is, however, general, and is not restricted by this assumption. The molecular layer is placed inside the DBR cavity at $y=0$. 

\begin{figure*}[h]
    \centering
    \includegraphics[width=0.7\textwidth]{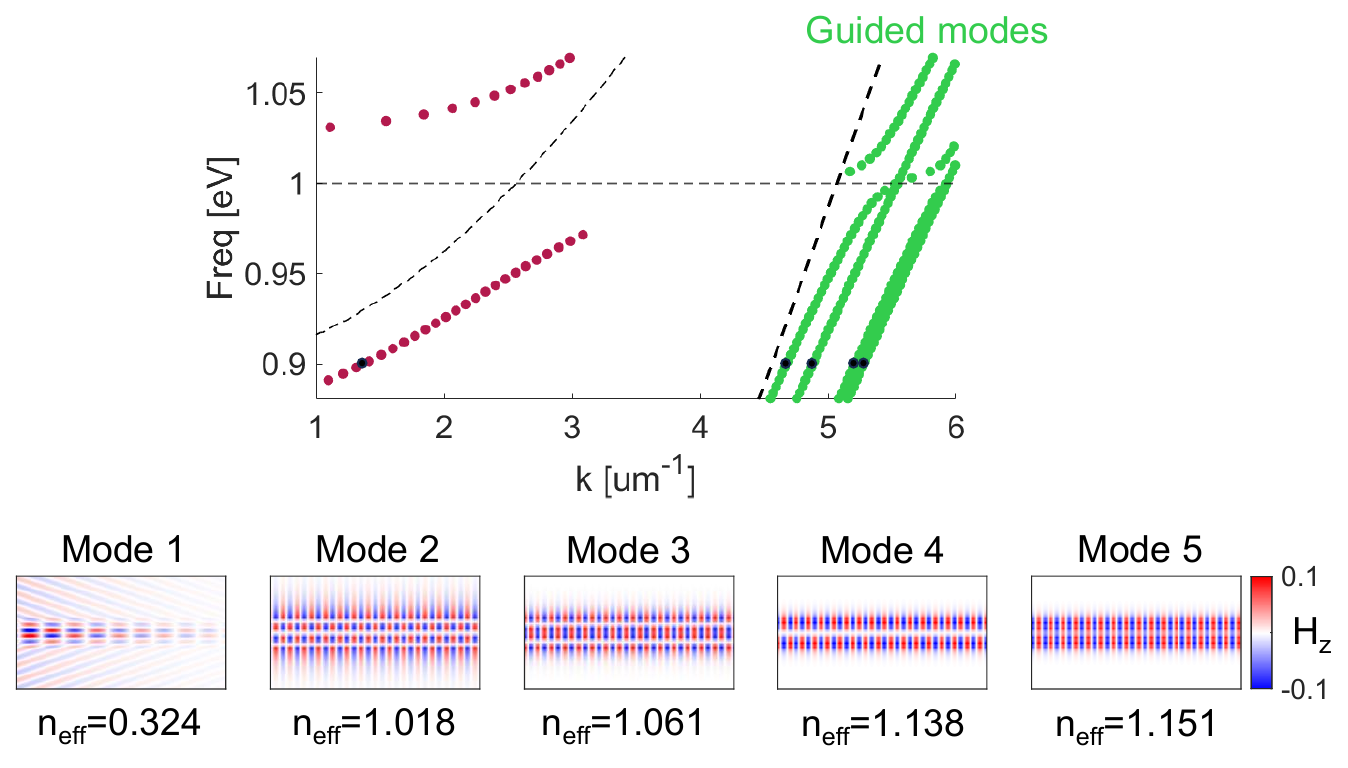}
    \caption{Eigenmodes supported by our emitter-cavity system. In the top panel, each obtained eigenmode is represented by a dot. The bottom panels visualize the $H_{z}$ field distributions of five modes located at $\omega = \omega_{c} = 0.9$ eV. }
    \label{fig_SI:eigenmodes}
\end{figure*}

By substituting Eq.~\eqref{eq:S_wg_mode_ansatz} into Eq.~\eqref{eq:S_TM_master}, we arrive at a one-dimensional eigenvalue problem
\begin{equation}
    \left[
    \varepsilon_{yy}
    \frac{\partial}{\partial y}
    \left( \varepsilon_{xx}^{-1} \frac{\partial}{\partial y} \right)
    +
    \varepsilon_{yy} k_{0}^{2}
    \right]
    \psi(y)
    =
    n_{\mathrm{eff}}^2 k_{0}^{2}
    \psi(y),
    \label{eq:S_wg_eigenproblem}
\end{equation}
where $n_{\mathrm{eff}} = \beta / k_{0}$ denotes the effective refractive index. 
% This is the same TM FDFD operator used in the inverse-design calculations, specialized here to a translationally invariant slab.
The above Eq.~\eqref{eq:S_wg_eigenproblem} is then discretized and treated as an eigenvalue problem at $\omega=\omega_c=0.9$~eV to obtain the mode profiles in Fig.~\ref{fig:fig3}(b). % entering $\varepsilon_{xx}$ as an effective Lorentzian sheet regularized by a small $\gamma_\phi=10^{-12}\omega_0$.
We introduce stretched-coordinate PMLs in $y$ direction as a boundary condition. 
Note that solving for the above eigenvalue problem leads to spurious modes that are localized inside the PML. 
These spurious modes are rejected and are not included in our analysis. 
We also reject strongly decaying modes with $\operatorname{Im}(n_{\mathrm{eff}})>1/(2\pi)$. 
The remaining solutions, as shown in Fig.~\ref{fig_SI:eigenmodes}, are classified into two categories: modes with $\operatorname{Re}(n_{\mathrm{eff}})<1$ are above the light line and can couple to the radiative channels in free-space, while modes with $\operatorname{Re}(n_{\mathrm{eff}})>1$ lie below the light line and do not couple to the radiative channels. 

% The guided-mode loss is thus several orders of magnitude smaller, which is why this channel supports low-loss propagation. 
% Their transverse profiles $\psi(y)=H_z(y)$ are cropped to the physical region and saved as the source and target profiles for the converters below.

\newpage
\section{Inverse design of the mode converters}
\label{sec:inverse_design}
Here we describe the procedure used to design the two mode converters shown in Fig.~\ref{fig:fig3}(c). 
The calculation of mode profiles has already been displayed in Sec.~\ref{sec:waveguide_mode}. 
The goal is to construct compact grating-like structures inside the DBR cavity that transform between the radiative mode (mode 1 in Fig.~\ref{fig_SI:eigenmodes}) and the low-loss guided mode (mode 2 in Fig.~\ref{fig_SI:eigenmodes}). 
Two converters are designed independently: converter~1 transforms the radiative mode into the guided mode, while converter~2 does the opposite. 

The optimization is performed with the help of a finite-difference frequency-domain (FDFD) solver, which solves for the magnetic field $H_z$. 
Specifically, after discretization the field vector $\bm h$ satisfies a linear equation $\bm A \bm h = \bm b_0$, 
where $\bm A$ denotes the following linear operator
\begin{equation}
    \bm A =
    \bm D_x^{b}\,\mathrm{diag}\!\left(\varepsilon_{yy}^{-1}\right)\bm D_x^{f}
    +
    \bm D_y^{b}\,\mathrm{diag}\!\left(\varepsilon_{xx}^{-1}\right)\bm D_y^{f}
    +
    k_0^2 \bm I,
    \label{eq:S_inv_operator}
\end{equation}
and $\bm b_0$ stands for the injected mode source. 
Here $\bm D_{x,y}^{f}$ and $\bm D_{x,y}^{b}$ are the matrices corresponding to the forward and backward finite-difference operators. 
Stretched-coordinate PMLs are introduced in both $x$ and $y$ directions, and Dirichlet boundary conditions are imposed on the outermost grid points. 
% the sheet currents are represented on one grid cell as
% \begin{equation}
%     J_y^{\mathrm{src}}(y)=\frac{H_z^{\mathrm{src}}(y)}{\Delta y},
%     \qquad
%     M_z^{\mathrm{src}}(y)=\frac{E_y^{\mathrm{src}}(y)}{\Delta y}.
% \end{equation}
% These currents form the source vector $\bm b_0$ in Eq.~\eqref{eq:S_inv_forward_system}.
We now denote the targeted mode profile at the output plane as $\bm h_{\mathrm{out}}$. 
The figure-of-merit (FoM) that we try to maximize is defined as
\begin{equation}
    \mathrm{FoM}=|\bm h_{\mathrm{out}}^{\dagger}\bm h|^2 =
    \left|\sum_y
    H_{\mathrm{out}}^*(x_{\mathrm{out}},y)\,
    H_z(x_{\mathrm{out}},y)\right|^2,
    \label{eq:S_inv_fom}
\end{equation}
which quantifies the overlap between the actual field $\bm h$ and the desired mode. 

\subsection{Parameterization of permittivity distribution}
The background structure is the same DBR cavity described in Sec.~\ref{sec:dbr_cavity}. 
Note that the high-index dielectric layers, as well as the molecular sheet, remain fixed and cannot be modified during the optimization process. 
Only the low-index layers are designable. 
% For $N_{\mathrm{DBR}}=3$, this gives $2(N_{\mathrm{DBR}}+1)=8$ designable layers. 
In the $l$-th designable layer, we start from a one-dimensional pattern $p_l(x)$, which controls the permittivity distribution along the $x$ direction. 
This pattern is tiled along the $y$ direction, so the final structure is a fabricable layered grating rather than a free-form 2D pattern. 
At each iteration, the latent variable $p_l(x)$ has to be mapped to a permittivity distribution. %through a filter-and-projection pipeline.
A sigmoid function is first applied:
\begin{equation}
    \rho_{\mathrm{raw}}(x,l)
    =
    \frac{1}{1+\exp[-p_l(x)]}.
\end{equation}
A conic filter along $x$ is then applied, which helps control the minimum feature size:
\begin{equation}
    \tilde{\rho}(x,l)
    =
    \frac{[K \circledast \rho_{\mathrm{raw}}](x,l)}
         {[K \circledast 1](x,l)},
    \qquad
    K(r)=\max(0,r_{\mathrm{filter}}-|r|),
\end{equation}
here $\circledast$ represents 1D convolution. 
The filtered density $\tilde{\rho}(x,l)$ is then projected toward a binary value using
\begin{equation}
    \rho(x,l)
    =
    \frac{
    \tanh(\beta\eta)
    +
    \tanh[\beta(\tilde{\rho}(x,l)-\eta)]
    }
    {
    \tanh(\beta\eta)
    +
    \tanh[\beta(1-\eta)]
    },
    \label{eq:S_inv_tanh_projection}
\end{equation}
where $\beta$ denotes the projection parameter which controls the binarization. The threshold is fixed as $\eta=0.5$.
Finally, the material permittivity can be obtained through interpolation:
\begin{equation}
    \varepsilon^{-1}
    =
    \frac{1}{\varepsilon_2}
    +
    \left(
        \frac{1}{\varepsilon_1}
        -
        \frac{1}{\varepsilon_2}
    \right)\rho,
    \label{eq:S_inv_material_interp}
\end{equation}
here $\varepsilon_{1} = n_{1}^{2}$ and $\varepsilon_{2} = n_{2}^{2}$. 
$\rho=0$ means that the gap is filled with high-index material. 

\subsection{Calculating gradient using adjoint method}
For the linear system $\bm A \bm h = \bm b_0$, the Wirtinger derivative of the FoM with respect to the conjugate field is
\begin{equation}
    \frac{\partial \mathrm{FoM}}{\partial \bm h^*}
    =
    \bm h_{\mathrm{out}}\, (\bm h_{\mathrm{out}}^{\dagger}\bm h) .
\end{equation}
The adjoint field is therefore obtained from one additional linear solve,
\begin{equation}
    \bm A^\dagger \bm h_{\mathrm{adj}}
    =
    -\bm h_{\mathrm{out}}\, (\bm h_{\mathrm{out}}^{\dagger}\bm h) .
    \label{eq:S_inv_adjoint}
\end{equation}
Each optimization iteration therefore requires one forward solve and one adjoint solve.
% Because the operator depends on the design only through diagonal factors of $\varepsilon_{xx}^{-1}$ and $\varepsilon_{yy}^{-1}$, 
The sensitivity with respect to inverse permittivity $\varepsilon^{-1}$ can be derived as
\begin{equation}
    \frac{\partial \mathrm{FoM}}{\partial \varepsilon^{-1}}
    =
    2\,\mathrm{Re}
    \left[
        v_x^*u_x+v_y^*u_y
    \right],
    \label{eq:S_inv_epsinv_grad}
\end{equation}
where
\begin{equation}
    u_x=\bm D_x^f\bm h,\quad
    u_y=\bm D_y^f\bm h,\quad
    v_x=(\bm D_x^b)^\dagger\bm h_{\mathrm{adj}},\quad
    v_y=(\bm D_y^b)^\dagger\bm h_{\mathrm{adj}} .
\end{equation}
% The $x$-flux term corresponds to $\varepsilon_{yy}^{-1}$ and the $y$-flux term corresponds to $\varepsilon_{xx}^{-1}$; inside the design region the two tensor components are varied together, so both contributions are accumulated into the same design gradient.
Finally, the gradient is propagated back through the parameterization in reverse order. 
% First, the cellwise gradient is summed over all grid rows belonging to each designable layer.  The binarization gradient from Eq.~\eqref{eq:S_inv_objective} is added, followed by the derivative of the tanh projection in Eq.~\eqref{eq:S_inv_tanh_projection}, the transpose of the conic filter, and the sigmoid derivative $\rho_{\mathrm{raw}}(1-\rho_{\mathrm{raw}})$. 
The resulting gradient with respect to $p_l(x)$ is passed to the optimizer for gradient descent. 

The objective $J$ we try to optimize includes both the FoM and a binarization penalty term:
\begin{equation}
    J=\mathrm{FoM}- \lambda_{\mathrm{bin}}
    \sum_{\mathrm{cells}}\rho(1-\rho),
    \label{eq:S_inv_objective}
\end{equation}
where $\lambda_{\mathrm{bin}}$ is increased during the optimization.
The penalty term pushes $\rho$ towards $0$ or $1$. 

\subsection{Numerical settings}

We carry out the inverse design focusing on a single wavelength $\lambda_c=1377.6$~nm. 
The grid resolution is fixed as $\Delta x=17$~nm. The design region is $8~\mu$m long in $x$ direction. 
During the optimization process, we run gradient descent (based on Adam optimizer) for $600$ steps. 
The learning rate starts from $0.1$ and is decayed by a factor $0.9958$ per iteration. 
To control the minimum feature size, the radius of the applied conic filter is chosen as $r_{\mathrm{filter}}=400$~nm. 
The projection parameter is ramped from $\beta=1$ to $\beta=150$, to ensure that the resulting permittivity profile is binary. 
The binarization weight $\lambda_{\mathrm{bin}}$ is kept at zero for the first $90$ steps, and then increased linearly to $0.15$. 
During the final $150$ steps, the best-performing device is recorded.
The final device is obtained by thresholding the profile at $\rho=0.5$, before being tested using FDTD. 

The thresholded grating structures are the shaded regions shown in Fig.~\ref{fig:fig3}(c).  The same exported binary permittivity, after decomposition into rectangular features, is used in the long-distance propagation simulations of Fig.~\ref{fig:fig4}.  Thus the inverse-design calculation supplies only the passive photonic converters; the subsequent nonlinear pump--probe dynamics are evaluated with the full Maxwell--Liouville FDTD framework described in the main text. 

\begin{figure*}[t]
    \centering
    \includegraphics[width=0.9\textwidth]{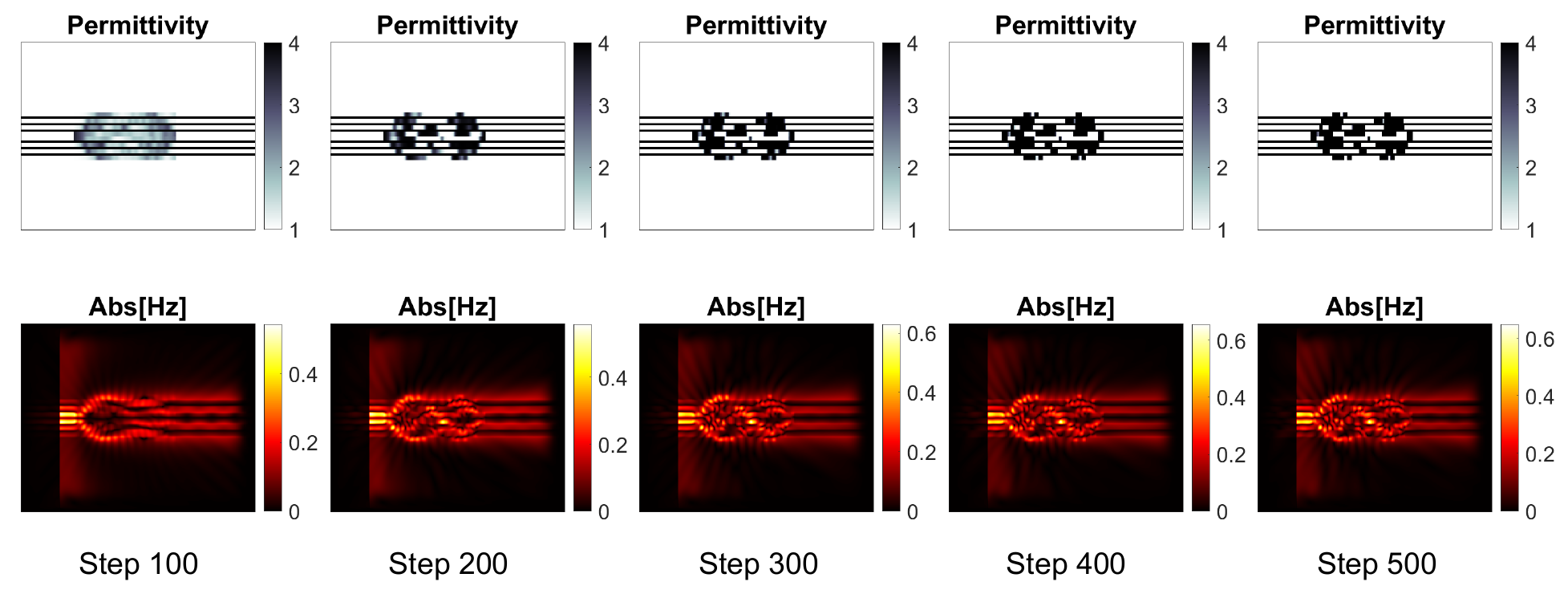}
    \caption{Optimization of the mode converter. The top row shows how the permittivity distribution evolves as the number of steps increases. The bottom row shows the corresponding $|H_{z}|$ field distributions. 
    }
    \label{fig_SI:optimization}
\end{figure*}

\newpage
\begin{figure*}[t]
    \centering
    \includegraphics[width=0.9\textwidth]{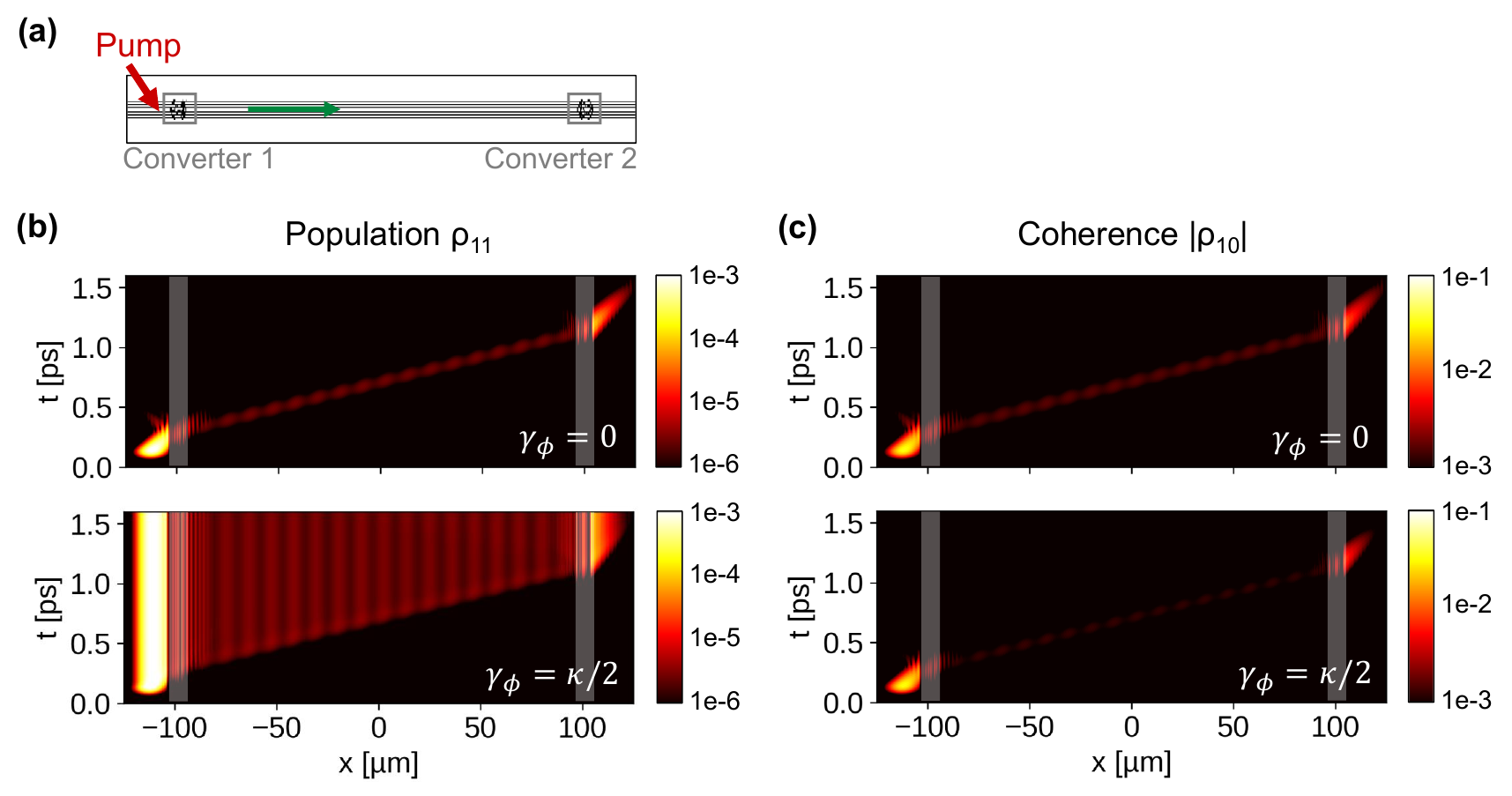}
    \caption{Long-range excitation transport enabled by the mode converters. 
    (a) Simulation setup. The distance between the two mode converters is set to $200~\mu$m. 
    (b) Heatmap of the population $\rho_{11}(x, t)$. When $\gamma_{\phi}=\kappa/2$, the heatmap shows the creation of static dark state population. 
    (c) Heatmap of the coherence $|\rho_{10}(x, t)|$. The coherence is only slightly affected by $
    \gamma_{\phi}$.  }
    \label{fig_SI:converter_coherence}
\end{figure*}

\newpage
\begin{figure*}[t]
    \centering
    \includegraphics[width=0.7\textwidth]{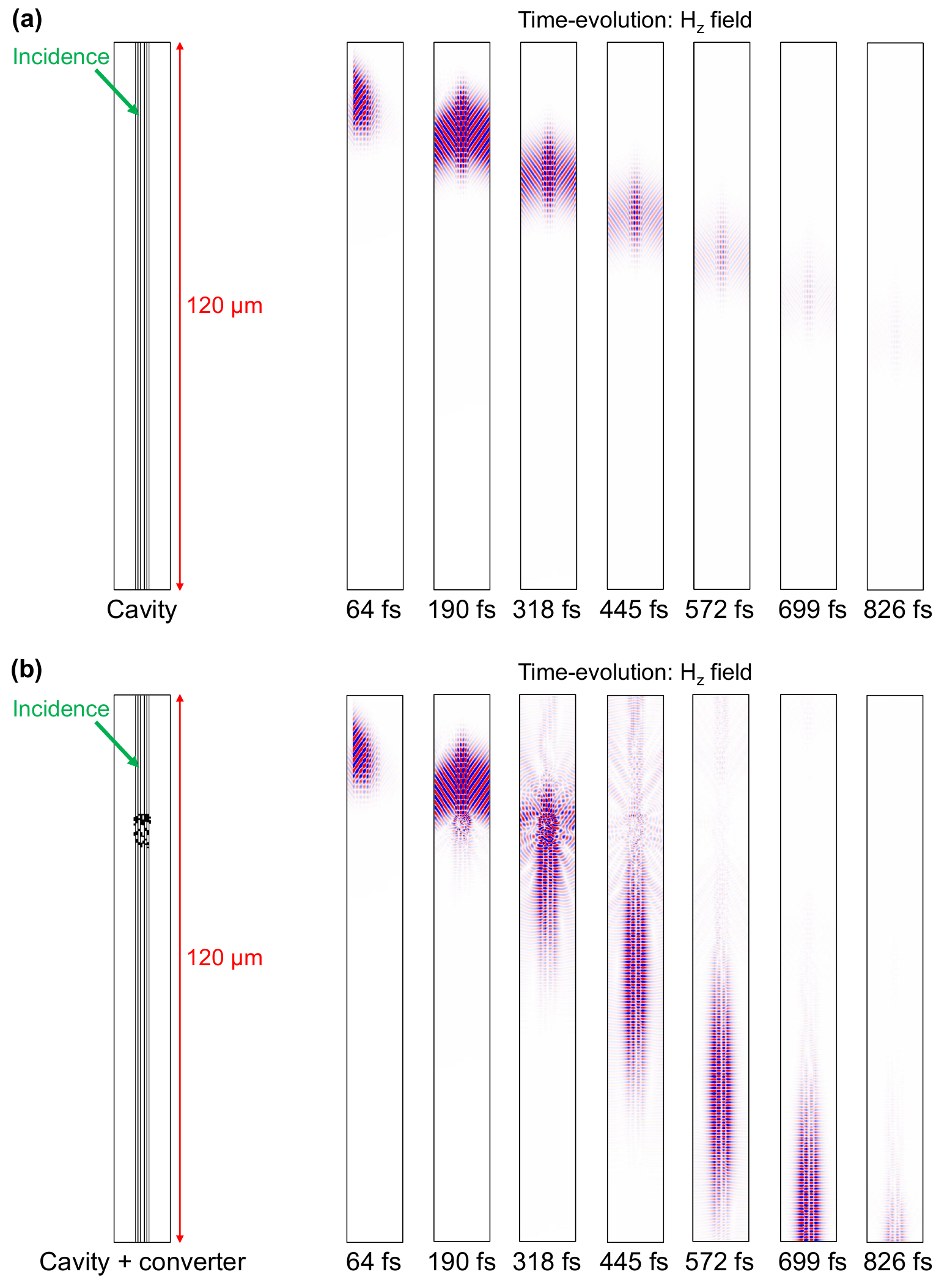}
    \caption{Visualization of magnetic field snapshots to illustrate how the mode converter enables long-range energy transport. All the field snapshots are obtained through FDTD simulation. 
    (a) Without the mode converter, the excited cavity polariton mode is radiative and decays quickly. 
    (b) With the mode converter, the polariton mode is converted into a guided mode which is protected from radiative losses.}
    \label{fig_SI:long_propagation_snapshot}
\end{figure*}

% \newpage
% \begin{figure*}[t]
%     \centering
%     \includegraphics[width=\textwidth]{figures_SI/weak_coupling.pdf}
%     \caption{Verifying that the long-range propagation still persists in the weak-coupling regime. Compared to Fig.~\ref{fig:fig4}, the dipole moment of the molecules is decreased by $100\times$. The black dashed lines mark $\omega_{0}=1$ eV and the light line. The blue dashed line indicates that the central frequency is $0.95$ eV.
%     (a) TMM results show the disappearance of the anti-crossing in the weak-coupling regime. 
%     (b) Population heatmap $\rho_{11}(x, t)$. Top panel: $\gamma_{\phi}=0$; bottom panel: $\gamma_{\phi}=\kappa/2$. The results show patterns similar to those in Fig.~\ref{fig:fig4} in the main text. The absolute molecular populations are much smaller, as expected from the reduced molecular dipole moment. 
%     (c) Maximum received molecular excitation $\max_{t} \rho_{11}(x, t)$ over time. With mode converters (blue curves) the excitation drops after converter 1, then rises after converter 2. As a comparison, without mode converters the excitation decays exponentially (red curves). }
%     \label{fig_SI:weak_coupling}
% \end{figure*}

\newpage
\begin{figure*}[t]
    \centering
    \includegraphics[width=0.7\textwidth]{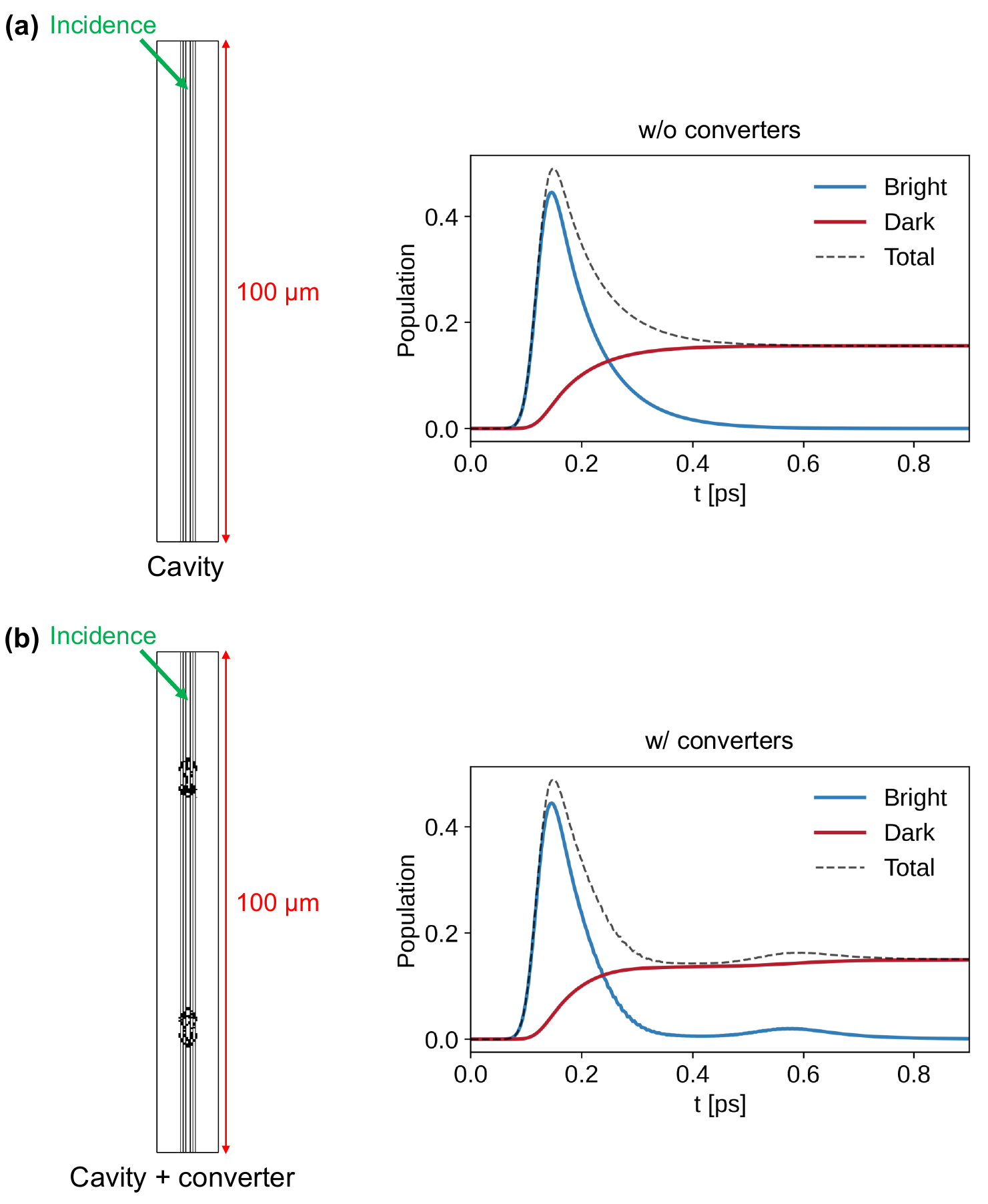}
    \caption{Visualization of bright and dark state populations. 
    Following Ref.~\cite{fowlerwright2026mapping}, we calculate the bright state population $n_{B}(t) = \sum_{j}| \rho_{10, j}(t)|^{2}$ and the dark state population $n_{D}(t)=\sum_{j}[\rho_{11, j}(t) - | \rho_{10, j}(t)|^{2}]$. 
    For both cases the pure dephasing rate is $\gamma_{\phi}=\kappa/10$. 
    (a) Without mode converter, $n_{B}(t)$ decays monotonically. Part of $n_{B}$ is transferred into static dark state population $n_{D}$ due to $\gamma_{\phi}>0$. 
    (b) With mode converters, $n_{B}(t)$ first decays, then rises and exhibits a recurrence peak at $t=0.57$ ps. This is because the energy of the injected polariton mode is transferred to the photon-like guided mode, then transferred back by the second converter. 
    }
    \label{fig_SI:population}
\end{figure*}

% \newpage
% \begin{figure*}[t]
%     \centering
%     \includegraphics[width=0.7\textwidth]{figures_SI/Tdiff_tau.pdf}
%     \caption{Differential transmission spectra $\Delta T(x, \omega)$ for different time delays $\tau$. The simulation setup is similar to that of Fig.~\ref{fig:fig4}(c) in the main text. The left column contains the results without pure dephasing ($\gamma_{\phi}=0$), while the right column contains the results for $\gamma_{\phi}=\kappa/2$. }
%     \label{fig_SI:Tdiff_tau}
% \end{figure*}

\newpage
\begin{figure*}[t]
    \centering
    \includegraphics[width=0.5\textwidth]{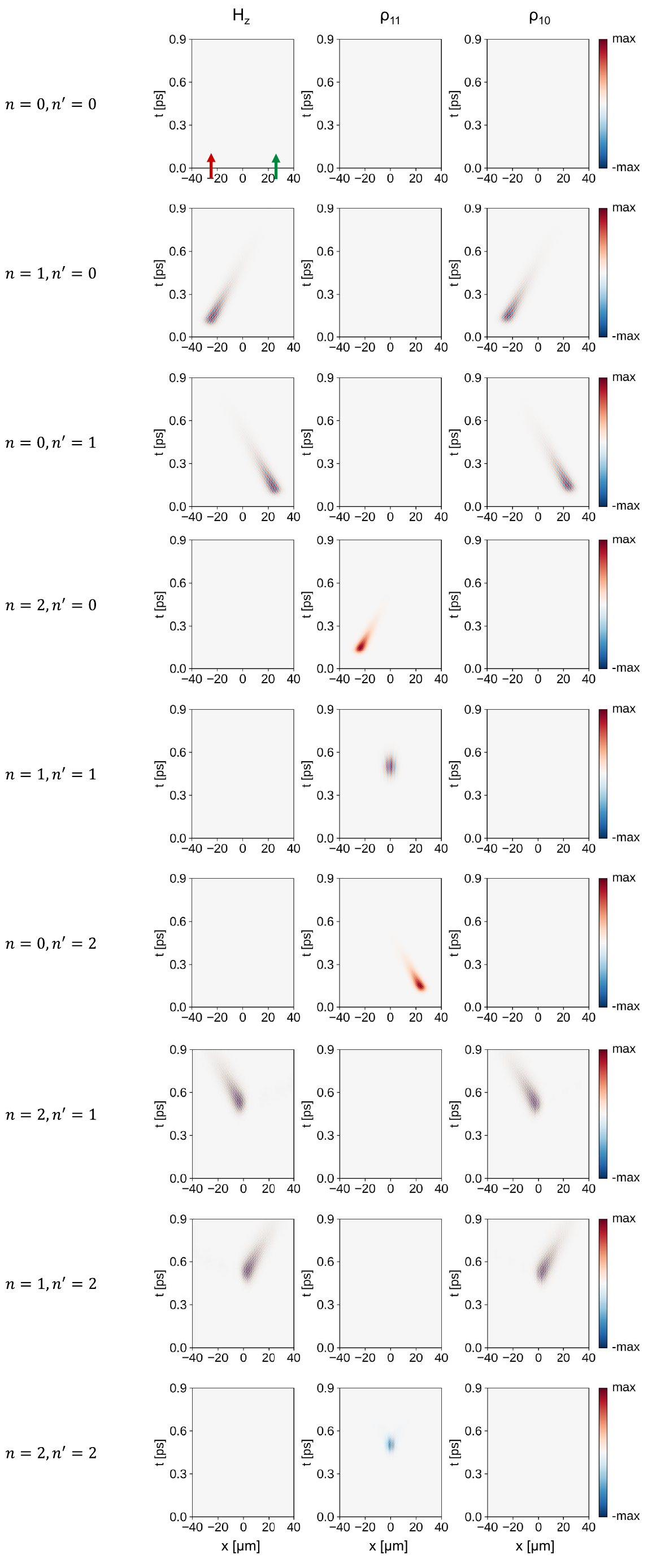}
    \caption{Recorded signals for different $(n)( n')$ orders. 
    The signals are obtained via FDTD simulation, using the simple cavity structure introduced in Fig.~\ref{fig:fig2}. 
    The red (green) arrow shows the center of the incident pump (probe) pulse. The corresponding in-plane wave vectors are $k_{p} = -k_{p'} = \pi/2~\mu$m$^{-1}$. 
    All fields are recorded at the $y=0$ central plane inside the DBR cavity. 
    The profiles are consistent with the equations introduced in Sec.~\ref{sec:perturbation_expansion}: if $n+n'$ is even, both the electromagnetic fields and the coherence $\rho^{(n)(n')}_{10}$ remain zero; if $n+n'$ is odd, the population $\rho^{(n)(n')}_{11}$ remains zero. 
     }
    \label{fig_SI:signal_orders}
\end{figure*}

\end{document}